\documentclass[twocolumn, deluxetables]{aastex631}
\usepackage{amsmath}
\usepackage[caption=false]{subfig}
\usepackage{multirow}
\usepackage{graphicx}
\usepackage{xcolor}
\usepackage[T1]{fontenc} 
\usepackage{enumitem}
\usepackage{comment}
\usepackage{rotating}
\usepackage{CJKutf8}

\shorttitle{NGC4151 Winds: When, Where and How}
\shortauthors{Xiang et al.}

\graphicspath{{./}{figures/}}
\begin{document}

\title{XRISM Spectroscopy of Variable Accretion-driven Disk Winds in  NGC 4151:\\ When, Where, and How Fast Outflows are Launched}

\author[0000-0002-7129-4654]{Xin Xiang
\begin{CJK*}{UTF8}{gbsn}
(项辛)
\end{CJK*}}
\affiliation{Department of Astronomy,
The University of Michigan, 1085 South University Avenue, Ann Arbor, Michigan,48103, USA}
\email{xinxiang@umich.edu}

\author[0000-0003-2869-7682]{Jon M. Miller}
\affiliation{Department of Astronomy,
The University of Michigan, 1085 South University Avenue, Ann Arbor, Michigan,48103, USA}

\author[0000-0002-4992-4664]{Missagh Mehdipour}
\affiliation{Department of Astronomy,
The University of Michigan, 1085 South University Avenue, Ann Arbor, Michigan,48103, USA}

\author[0000-0001-9735-4873]{Ehud Behar}
\affiliation{Department of Physics, Technion, Technion City, Haifa 3200003, Israel}

\author[0000-0002-0167-2453]{W. N. Brandt}
\affiliation{Department of Astronomy \& Astrophysics and the Institute for Gravitation and the Cosmos, The Pennsylvania State University, 525 Davey Lab, University Park, PA}

\author[0000-0003-2663-1954]{Laura Brenneman}
\affiliation{Center for Astrophysics | Harvard-Smithsonian, MA 02138, USA}

\author[0009-0006-4968-7108]{Luigi Gallo}
\affiliation{Department of Astronomy and Physics, Saint Mary's University, Nova Scotia B3H 3C3, Canada}

\author[0000-0002-0273-218X]{Elias Kammoun}
\affiliation{Cahill Center for Astrophysics, California Institute of Technology, 1216 East California Boulevard, Pasadena, CA 91125, USA}

\author[0000-0003-4511-8427]{Peter Kosec}
\affiliation{Center for Astrophysics | Harvard-Smithsonian, MA 02138, USA}
%\email{peter.kosec@cfa.harvard.edu}

\author[0000-0001-9911-7038]{Liyi Gu}
\affiliation{SRON Netherlands Institute for Space Research, Leiden, The Netherlands} %10
%\email{l.gu@sron.nl}

\author[0000-0002-3687-6552]{Doyee Byun 
\begin{CJK}{UTF8}{mj}
(변도의)
\end{CJK}}
\affiliation{Department of Astronomy,
The University of Michigan, 1085 South University Avenue, Ann Arbor, Michigan,48103, USA}

\author[0000-0002-7962-5446]{Richard Mushotzky}
\affiliation{Department of Astronomy, University of Maryland, College Park, MD 20742, USA}

\author[0000-0002-8108-9179]{Stephane Paltani}
\affiliation{Department of Astronomy, University of Geneva, Versoix CH-1290, Switzerland}

\author[0000-0001-8470-749X]{Elisa Costantini}
\affiliation{SRON Space Research Organization Netherlands, Niels Bohrweg 4, 2333CA Leiden, The Netherlands} 
\affiliation{Anton Pannekoek Institute for Astronomy, University of Amsterdam, Science Park 904, NL-1098 XH Amsterdam, The Netherlands}
%\email{E.Costantini@sron.nl}

\author[0000-0002-0572-9613]{Abderahmen Zoghbi}
\affiliation{Department of Astronomy, University of Maryland, College Park, MD 20742, USA}
\affiliation{HEASARC, Code 6601, NASA/GSFC, Greenbelt, MD 20771, USA}
\affiliation{CRESST II, NASA Goddard Space Flight Center, Greenbelt, MD 20771, USA}

%\collaboration{6}{(Collaborators)}

\begin{abstract}
X-ray observations probe the inner accretion flow within active galactic nuclei, revealing the highest gas column densities and fastest winds.  The most diverse winds yet revealed with the Resolve calorimeter spectrometer aboard XRISM are found in NGC~4151, a nearby Seyfert-1 AGN that may qualify as a ``changing-look'' source (CLAGN).  Herein, we report on wind variability in 14 XRISM observations of NGC~4151, summing to 0.9~Ms of exposure over a period of 395 days.  We examined the dependence of key wind parameters on hardness and intensity selections, and as a function of time relative to flaring and dip events. The results suggest a globally organized but locally complex wind structure. Slow ``warm absorber'' components (WAs; $v_{\rm{out}} \sim 100-1000~\rm{km~s^{-1}}$) are always observed and likely represent failed winds at radius of $10^4 - 10^5 GM/c^2$, within the inner wall of the torus.  In contrast, ``very fast'' and ``ultra-fast'' outflows (VFOs and UFOs; $v_{\rm{out}} \sim 10^3-10^4~\rm{km~s^{-1}}$, $v_{\rm{out}} \sim 0.033-0.33~c$) are strongest 10~ks after the peak of flares, and during periods with low flux. Ten kiloseconds is among the shortest flare--wind response timescales reported in an AGN, suggesting that the winds are observed close to the launching site. The absorption measure distribution (AMD) and the large outflow momentum rates suggest that the high-velocity flows visible in the Fe~K band are magnetically driven, while locally clumpy, likely owing to radiation pressure; one or both of these mechanisms may be enhanced following a flare and most visible during low-flux windows.
\end{abstract}


\section{Introduction} \label{sec:intro}
Accretion onto supermassive black holes (SMBHs) \citep{Soltan_1982, Lynden-Bell_1969} can drive outflows that carry mass, momentum, and energy into their host galaxies \citep{King_Pounds_2015}. Powerful, ionized winds are thought to play an important role in active galactic nucleus (AGN) feedback, potentially regulating star formation and black hole growth (e.g. \citealt{Di_2005, Hopkins_2010}).  This regulation may explain scaling relations between galactic properties and black hole masses \citep{Kormendy_Ho_2013, Greene_2020}.  Simulations suggest that wind feedback can halt star formation in host bulges when the kinetic power in winds exceeds 0.5--5\% of the Eddington limit of the black holes (e.g. \citealt{Di_2005, Hopkins_2010}).

X-ray spectroscopy can probe the most ionized black hole winds, wherein even Fe is found in He-like and H-like charge states.  If the ionization of such wind components is due to their proximity to the black hole, these winds should have especially high velocities and show rapid variability.  The Resolve calorimeter aboard XRISM (\citealt{Ishisaki_2022}, \citealt{Tashiro_2025}) has clearly revealed variable ``ultra-fast'' outflows (or, UFOs) in NGC~4151, NGC 3783, and PDS~456 (e.g., \citealt{Xiang_2025}, \citealt{Gu_2025}, \citealt{XRISMPDS456_2025}), among other AGN with emerging evidence of fast winds.

NGC~4151 is an especially interesting case.  Unlike PDS~456 and other quasars, Resolve spectra reveal the full array of outflow types and velocities, even in just the Fe~K band.  These wind types include ``warm absorber'' flows that were revealed in low-energy X-rays with Chandra and XMM-Newton (WAs; $v \sim 10^2$--$10^3$ km s$^{-1}$; e.g., \citealt{Couto_2016}, \citealt{XRISM_NGC4151_2024}, \citealt{Xiang_2025}, \citealt{Miller_2026}), ``very fast outflows'' (VFOs: $v \sim 10^3$--$10^4$ km s$^{-1}$, \citealt{Xiang_2025}, \citealt{Miller_2026}; and/or ``eclipsing winds'', \citealt{Longinotti_2019} \citealt{Mehdipour_2017}), and ultra-fast outflows (UFOs; $v \sim 0.033$--$0.33c$, \citealt{Tombesi_2010}, \citealt{Xiang_2025}, \citealt{Miller_2026}).   The simultaneous production of these winds in NGC 4151 may suggest a stratified but connected multi-phase structure with common driving mechanisms, but more study is required (for a broad review of AGN winds and emission lines in the X-ray bands, see, e.g., \citealt{Gallo_2023}).

NGC~4151 (z = 0.0033, $M_{\rm{BH}} = 3.4^{+0.4}_{-0.4} \times 10^7 M_{\odot}$ \citealt{Bentz_2015}) is one of the brightest and best-studied nearby Seyfert galaxies, making it an ideal laboratory for studying the physics of AGN outflows. It is sometimes regarded as a ``changing-look'' AGN, or CLAGN, owing to variations in its optical lines, as well as its X-ray flux and complex internal obscuration \citep{Ives_1976, Ulrich_1997, Edelson_2017, Miller_2018}.  NGC~4151 is formally ``radio quiet,'' but it exhibits collimated outflows with sub--relativistic speeds \citep{Ulvestad_2005}.  The narrow line region (NLR) in NGC~4151 has been resolved as a biconical outflow in the optical, UV, and X-ray bands, and exhibits the typical signatures of photoionizaton (e.g., \citealt{Das_2005}, \citealt{Wang_2011}). 

The 14 observations of NGC~4151 obtained with XRISM (so far) represent a unique opportunity to understand how and when the fastest outflows in sub-Eddington AGN are launched, and to better evaluate their ability to reshape the host galaxy.  The sub-Eddington regime may differ substantially from the near-Eddington and super-Eddington regimes.  For instance, the UFO absorption lines in NGC~4151 appear to be quite broad, with $\sigma \sim 5\times 10^{3}~{\rm km}~{\rm s}^{-1}$ \citep{Xiang_2025}, whereas those observed in PDS~456 are narrower $\sigma \sim 1.9\times 10^{3}~{\rm km}~{\rm s}^{-1}$ \citep{XRISMPDS456_2025} despite bulk velocities that are 2--3 times higher.  The former case is qualitatively more consistent with continual driving, via radiation pressure or magnetic pressure, while the latter is more consistent with ballistic ejection.  In XRISM observations of the Seyfert-1 AGN NGC~3783, however, a UFO is revealed approximately 50~ks after a large soft X-ray flare \citep{Gu_2025}, potentially powered by magnetic reconnection (as per a solar coronal mass ejection \citep{Vardi_2026}).  Together, these suggest that UFOs may not arise from a single launching mechanism but instead may reflect different combinations of radiation pressure, magnetic driving, and impulsive coronal activity across accretion regimes.

In \cite{Xiang_2025}, hereafter Paper I, we fit self-consistent photoionization models to the five XRISM observations of NGC~4151 that were obtained during its ``Performance and Verification'' phase.  Across these observations, we found significant variations in the properties of the fast outflows.  Herein, we consider all 14 observations of NGC~4151, but whereas Paper I only considered entire observations as indivisible units, this analysis examines variations in flux and spectral properties within individual observations, and across multiple observations.  In short, physical considerations related to flare and dip response times and changes in ionizing flux dictate how spectra are grouped and fit in this work. This work is organized around three related questions: \textit{when}, \textit{where}, and \textit{how} are the winds in NGC~4151 launched? Section \ref{sec:data} describes the data reduction process for the XRISM observations. In Section \ref{sec:results}, we outline the data analysis methods, the spectral fitting methods, and present the fitting results. We further explore the implications of these findings in Section \ref{sec:discussion}, followed by Section \ref{sec:summary}, which summarizes our conclusions.

\section{Observation and Data Reduction} \label{sec:data}
We utilized all 14 XRISM/Resolve observations \citep{Tashiro_2025, Ishisaki_2022} during the Performance \& Verification phase and Cycle 1 (Obs IDs: 000125000, 000137000, 300047020, 300047030, 300047040, 201076010, 201076020, 201076030, 201076040, 201076050, 201076060, 201076070, 201076080, 201076090) on NGC 4151.  The summed spectrum from these observations is reported in \cite{Miller_2026}.  To ensure consistency with our prior work, our data reduction procedures closely followed those used in  Paper I.  Specifically, we extracted spectra (hi-res primary events `Hp') and lightcurves following XRISM Quick Start Guide v 2.3 and HEASOFT version 6.34 using the XRISM CalDB 11, and made response files using the standard \texttt{rslmkrmf} and \texttt{xaarfgen} tools.  We excluded data from pixels 12 (calibration), and 11 and 27 (these pixels sometimes produce anomalous data).  The FTOOL \texttt{maketime} was used to generate good time filters for a total of seven distinct windows based on time intervals and/or a combination of flux and hardness (detailed in \S \ref{sec:results}).  Distinct spectra from these windows were created using FTOOLS \texttt{mathpha} and \texttt{ftaddrmf}. 

\section{Data Analysis and Results} \label{sec:results}
In this section, we detail the selection criteria used to create seven spectral windows, the results derived from fitting each window with self-consistent photoionization models, and the emergent characteristics of the wind components.  The spectral analysis procedure closely followed the processes adopted in Paper I.  Briefly, the spectra were analyzed using SPEX version 3.08.01 \citep{Kaastra_1996}, minimizing a Cash statistic (C-stat; \citealt{Cash_1979}).  Within SPEX, the photoionization modeling package ``\texttt{pion}" \citep{Miller_2015, Mehdipour_2016} was used to fit the absorption and emission features within the data. We adopted the default SPEX cosmology for luminosity and distance-dependent quantities, with \(H_0=70~{\rm km~s^{-1}~Mpc^{-1}}\), \(\Omega_{\rm m}=0.3\), \(\Omega_{\Lambda}=0.7\), and \(\Omega_{\rm r}=0\). All of the errors reported in this work reflect the value of a parameter on the boundary of its $1\sigma$ confidence interval.  The significance of each wind component was estimated by calculating the Akaike information criterion (AIC; \citealt{Akaike_1974}) and the Detection Significance (DS); see Appendix \ref{adx:DS} for details.

\subsection{Light Curves, Hardness Intensity Diagram, and Event Type Selections} \label{subsec:LC}

\begin{figure*}
\centering
\rotatebox{90}{%
  \begin{minipage}{\textheight} % rotated width becomes \textheight
    \centering
    \includegraphics[width=\linewidth]{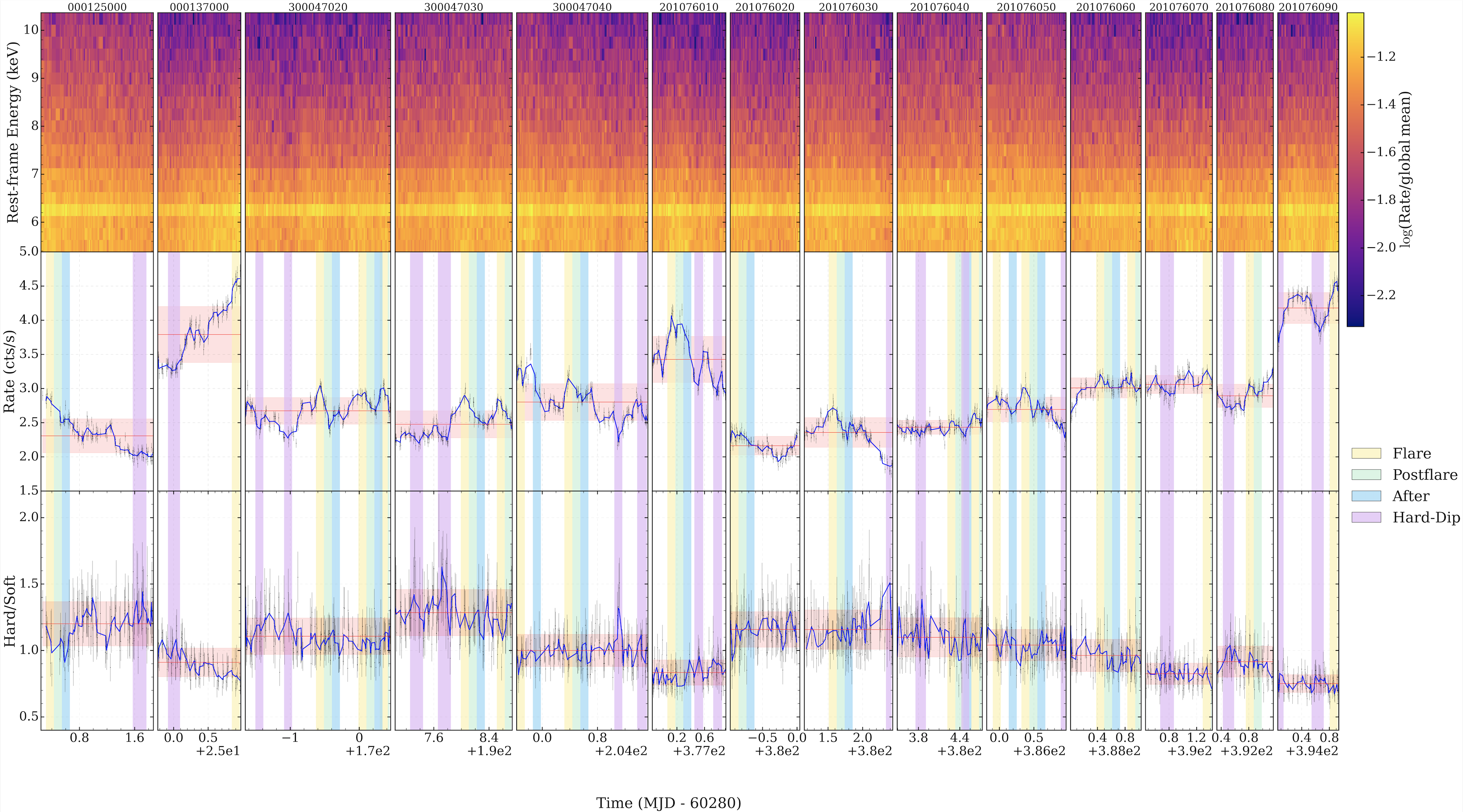}
    \caption{Top Panels: Trailed spectra.  Middle Panels: full band (2.4--17.4 keV) lightcurves. Bottom Panels: hardness ratios (8.8--11.8~keV / 3.0--4.0~keV)}
    \label{fig:LC}
  \end{minipage}%
}
\end{figure*}

Figure \ref{fig:LC} shows three light curves from the 14 Resolve observations of NGC~4151.  In each light curve, the time bin size is 520~s.  The top panels in Figure \ref{fig:LC} are trailed spectra, displaying the relative strength of different energy bins relative to a global mean.  The narrow Fe~K$_{\alpha}$ line at 6.40~keV is particularly prominent; variations in its strength are also visible.  The trailed spectrum also shows strong, rapid variability in the band above 8.8~keV, which is the band that ionizes Fe~XXV and Fe~XXVI.  The middle panels in Figure \ref{fig:LC} simply show the total count rate across the full Resolve band (2.4--17.4~keV).  The bottom panels in Figure \ref{fig:LC} show a hardness ratio, defined as the ratio count rate in the 8.8--11.8~keV band to the count rate in the 3.0--4.0~keV band (e.g. $H/S = CR_{8.8-11.8}/CR_{3.0-4.0}$).  The hard band is chosen to trace the flux that ionizes the most prominent charge states of iron in the wind absorption spectrum.  Importantly, this band avoids the strongest UFO absorption at 8.8~keV \citep{Xiang_2025}, so defining the response of UFOs relative to this band does not result in an accidental auto-correlation.  Within Figure \ref{fig:LC}, it is clear that NGC~4151 exhibits the ``softer when brighter'' behavior that is typical of Seyfert AGN, and previously indicated in long Chandra exposures \citep{Miller_2018}.  

Guided by the variations that are revealed in Figure \ref{fig:LC} and by the onset of a UFO after a flare in NGC~3783 \citep{Gu_2025}, we constructed spectra from four windows: (1) flares, (2) initial post-flare decay periods, (3) secondary post-flare decay periods, and (4) spectrally hard dip periods.  To separate significant variability in the full-band light curve and hardness curves from Poisson noise, we applied a Savitzky--Golay (SG) filter \citep{savitzky1964smoothing} with a window length of 11 time bins and a polynomial degree of 5; that is, each local 11-bin segment was smoothed by fitting a fifth-order polynomial. The spectral selecting windows were defined as follows:

\paragraph{Flaring intervals} Flares:
Local maxima in the smoothed full-band light curve were flagged as flares if their peak flux exceeded the mean level by at least one standard deviation.  Each flare interval was defined as a contiguous window surrounding the peak ($\pm$5 ks), truncated where necessary by the observation boundaries.

\paragraph{Initial post-flare intervals} Postflares:
For every flare interval, we defined the initial post-flare interval as the 10~ks period immediately following the end of the flare. A post-flare interval was retained only if the full-band count rate remained below the flare peak flux throughout the entire 10~ks window.  

\paragraph{Secondary post-flare intervals} Afters:
To explore a more delayed spectral/temporal response to flares, we also extracted a secondary interval of 10 ks extending from the end  of the initial post-flare interval.  Secondary intervals were only retained if the count rate within them remained below the peak flare rate.  Uniformly enforcing this criteria on the primary and secondary decay periods amounted to requiring clean data, where the effects of flares on wind production could be better determined.

\paragraph{Hard-dips intervals} Hard-Dips:
Dips were identified through a two-fold requirement involving both the full-band light curve and the hardness ratio.  First, we searched for local minima in the SG smoothed full-band light curve that fall at least one standard deviation below the mean level. Second, we required that these low-flux periods coincide with significant hardening of the spectrum, where the hardness values exceed the mean by at least one standard deviation. This selection isolates episodes wherein the source becomes simultaneously fainter and spectrally harder.

\begin{figure*}
    \centering
    \includegraphics[width=1.0\linewidth]{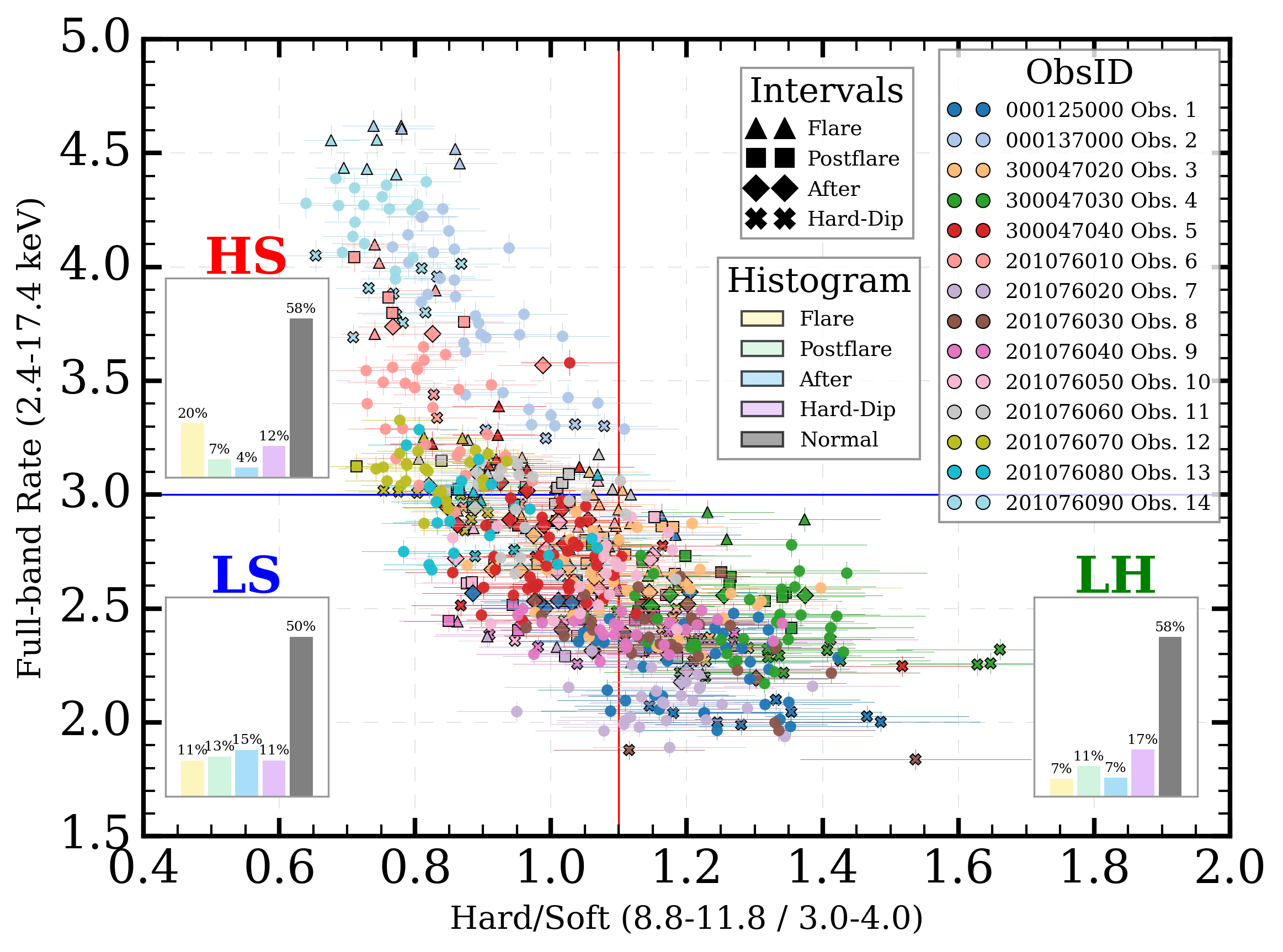}
    \caption{The Hardness-Intensity diagram: full band count rate versus Hard/Soft ratio. The red vertical and the blue horizontal lines mark the global mean of the hardness ratio and full-band count rates. The histogram shows the Time-bin Fraction of the local Event Types within each spectral region.}
    \label{fig:HRI}
\end{figure*}

In order to understand the response of wind components to basic changes in flux and spectral hardness (rather than sharp flares and dips), we constructed a hardness-intensity diagram (HID).  Figure \ref{fig:HRI} shows the count rate in the full Resolve band (2.4-17.4~keV), versus the hardness ratio used in Figure \ref{fig:LC} (8.8--11.8~keV / 3.0--4.0~keV).  In this plot, the intervals from the 14 observations trace a smooth ``banana'' figure, and bear out the ``softer when brighter'' behavior evident in Figure \ref{fig:LC}.  Although the points do not divide into separate groupings, we divided the data into three parts to begin exploring what might be a smooth evolutionary trend in wind parameters.  We defined a ``Low--Hard'' (LH) region for the intervals with a hardness ratio above the mean (spectrally harder than average) and count rates below the mean (fainter-than-average), and ``Low--Soft'' (LS) and ``High--Soft'' regions for the intervals with a hardness ratio below the mean and count rates below and above the mean, respectively.  

The time- and HID-selected spectra are designed to address \textit{when} different wind phases appear. This combined procedure provides a framework for tracking wind variations in response to both sharp and (potentially) more gradual changes in the ionizing flux from the central engine. Table \ref{tab:exposure} summarizes the total exposure accumulated in each window.  In Figure \ref{fig:HRI}, the ratio of the time-based windows represented in each hardness-intensity window is indicated.  The distribution of Flares, Postflares, Afters, and Hard-dips intervals varies across the hardness--intensity windows.  

\begin{table}[ht]
\centering
\caption{Total exposure accumulated in each variability-selected region.\label{tab:exposure}}
\begin{tabular}{cccc}
\hline
\hline
Event Type & Exposure (ks) & $L_{ion}$ & $F_{2-17\mathrm{keV}}$\\
\hline
Flares          & 97  & $3.95^{+0.13}_{-0.12}$ & $3.57^{+0.12}_{-0.11}$ \\
Postflares      & 77  & $3.58^{+0.16}_{-0.14}$ & $3.27^{+0.15}_{-0.13}$ \\
Afters          & 83  & $3.30^{+0.23}_{-0.13}$ & $3.18^{+0.22}_{-0.12}$ \\
Hard-Dips       & 115 & $3.45^{+0.10}_{-0.15}$ & $3.19^{+0.10}_{-0.14}$ \\
\hline
High--Soft (HS) & 243 & $3.52^{+0.06}_{-0.13}$ & $3.98^{+0.07}_{-0.15}$ \\
Low--Soft (LS)  & 307 & $3.49^{+0.25}_{-0.38}$ & $3.18^{+0.22}_{-0.35}$ \\
Low--Hard (LH)  & 310 & $3.17^{+0.13}_{-0.13}$ & $2.91^{+0.12}_{-0.12}$ \\

\hline
\end{tabular}
\tablecomments{The ionizing luminosities, $L_{\rm ion}$, are given in units of $10^{43}~\mathrm{erg~s^{-1}}$ and are calculated over 13.6 eV--13.6 keV. They represent the intrinsic luminosity emitted by the source, before dilution by absorption. The observed 2.4--17.4 keV fluxes are given in units of $10^{-10}~\mathrm{erg~cm^{-2}~s^{-1}}$. Both quantities are calculated from the best-fit models summarized in Table~\ref{table:parameters}. Errors are estimated from the uncertainty in the best-fit power-law normalization.}
\end{table}

\subsection{Modeling procedures} \label{subsec:model}
Spectral fits to each window included a continuum consisting of blackbody and a power-law components, fitted over the full pass band (2.4--17.4 keV).  These components loosely represent the emission from UV accretion disk and the hot X-ray corona. 
In all fits, the blackbody component ``bb" had a fixed peak temperature of $kT = 2 \times 16.8$ eV and an emitting area of $7.5\times10^{24}\,\mathrm{cm^2}$, consistent with spectral fits to Hubble Space Telescope Imaging Spectrograph (STIS) FUV spectra \citep{Kraemer_2006, Xiang_2025}. The power-law component was bent to zero at both high and low energies using two ``etau'' components (with cutoff energies of 0.0136 keV and 300 keV), ensuring a physically realistic spectral shape and ionizing luminosity. In addition to the power-law index $\Gamma$ and the flux normalization, we allowed the index break $\Delta \Gamma$ and break energy $E_0$ to vary, to account for a possible spectral hardening above $\sim 9$ keV due to potential unmodeled reflection features.

%The 'Mytorus' component
We model the Fe K$_\alpha$ and Fe K$_\beta$ emission lines using three ``Mytorus'' components \citep{Murphy_Yaqoob_2009}, modified by the ``spei'' blurring model \citep{Speith_1995} with a linked inclination, following the approach in \cite{XRISM_NGC4151_2024}. We also include the Ni K$_\alpha$ and Ni K$_\beta$ emission lines using two ``line'' models in SPEX, with their widths linked. The free parameters of ``Mytorus" components are the column density, inclination, and flux normalization. For the ``spei" component, the inner radius is free to vary. For the ``line" model, the optical depth and the FWHM of the Gaussian and Lorentzian components are free. The modeling of these neutral emission lines is not the focus of this work; they are included primarily to ensure that residuals in their spectral vicinity are not affected by unmodeled line flux. 

%The 'hot' component
As found in \citealt{Miller_2026}, the partial covering absorber is better described by a cool ionized absorber rather than a cold, neutral one, which shifts the Fe K edge to slightly higher energies than the neutral value (7.112 keV). We therefore adopt the same modeling approach here, using a ``hot'' component in which the gas temperature is allowed to vary freely, together with the equivalent neutral hydrogen column density ($N_{\rm H}$) and covering factor ($f_{\rm cov}$).

%The pion component, including emission and absorption
The remaining absorption and emission features are modeled using the photoionization components ``pion" \citep{Miller_2015, Mehdipour_2016}. The ``pion" components are arranged in layers with decreasing outflow velocities from inner to outer zones. During the fit, the ionizing luminosity incident on the outer zones is attenuated by the inner absorption zones along the line of sight. For each absorption component, the equivalent neutral hydrogen column density ($N_H~[\mathrm{cm^{-2}}]$), ionization parameter ($\xi~[\mathrm{erg~cm~s^{-1}}]$), line-of-sight outflow velocity in the source frame ($v_z~[\mathrm{km~s^{-1}}]$), and turbulent velocity ($\sigma_z~[\mathrm{km~s^{-1}}]$) are left free. The covering factor of the absorbers is fixed at $f_{\rm cov} = 0.5$, motivated by the population study of UFOs in AGNs \citep{Tombesi_2010}. It provides a uniform baseline for constraining the absorber column densities in consistent with Paper~1. Elemental abundances are fixed at solar values for all fits. For zones requiring substantial re-emission from the gas, we include the corresponding emission ``pion" component with their column density and ionization parameter linked to the counterpart absorptions. The emission covering factor ($\Omega_{\rm emis}$) and the emission velocity are allowed to vary.

The fits were performed iteratively over the broad band of 2.4--17.4 keV to constrain the continuum emission and the soft partial covering absorber, and more focused fits over the narrow band of 5.4--10.4 keV. Spectra were grouped with a binning factor of 15 between 5.4 keV and 7.4 keV, and a factor of 45 elsewhere. For the Resolve response, these correspond to spectral bin widths of 7.5~eV and 22.5~eV, respectively. We identify up to five statistically significant absorption zones. Not every spectrum requires all zones; nevertheless, we apply a consistent model framework to all spectra. For zones not required by the data (where $N_H$ regressed to zero), we estimate limits on $N_H$ by fixing the other wind parameters to their mean values and allowing the power-law normalization to vary. These components are not included in the final model since their significance is only $\le 1\sigma$. In addition to these absorption and re-emission zones, we detected transient blueshifted emission in some spectra. 

The best-fit parameters are summarized in Table~\ref{table:parameters}. Cash statistics ($C$) and the degrees of freedom (dof) are reported for the 5.4--10.4 keV band. The phenomenological SPEX form of the model written in the order from source to us for the $pion$ absorbers is

$((bb + (pow + 3 \times (mytorus * spei)) * etau_{low} * etau_{hi}) * line_{1} * line_{2} *  pion_{\#5} * pion_{\#4} * pion_{\#3} * pion_{\#2} * pion_{\#1} + vgau*pion^{emis\#2} + vgau*pion^{emis\#4} + vgau*pion^{blue-emis}) * hot * reds$

\begin{figure*}
    \centering
    \includegraphics[width=1.0\linewidth]{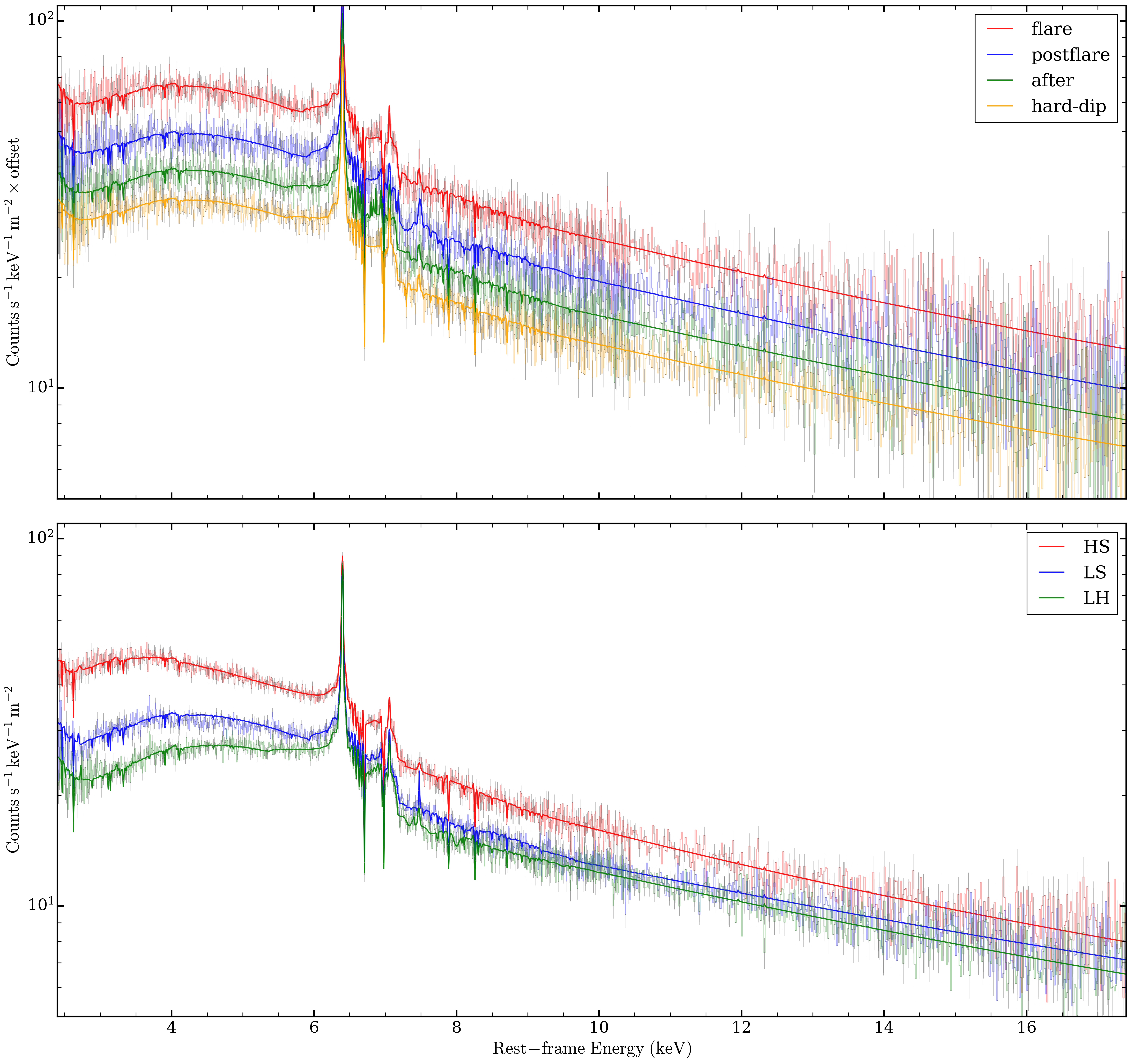}
    \caption{Upper panel: Spectra corresponding to time-selected phases (Flare, Postflare, After, and Hard-dip) with their best-fit models (Table \ref{table:parameters}), fitted over 2.4–17.4 keV. Lower panel: Spectra grouped by global spectral state and brightness: High-Soft (HS), Low-Soft (LS), and Low-Hard (LH). Data are binned with a factor of 15 below 10.4 keV and 35 above 10.4 keV for visual purposes.}
    \label{fig:brd}
\end{figure*}

\begin{figure*}
    \centering
    \includegraphics[width=0.95\linewidth]{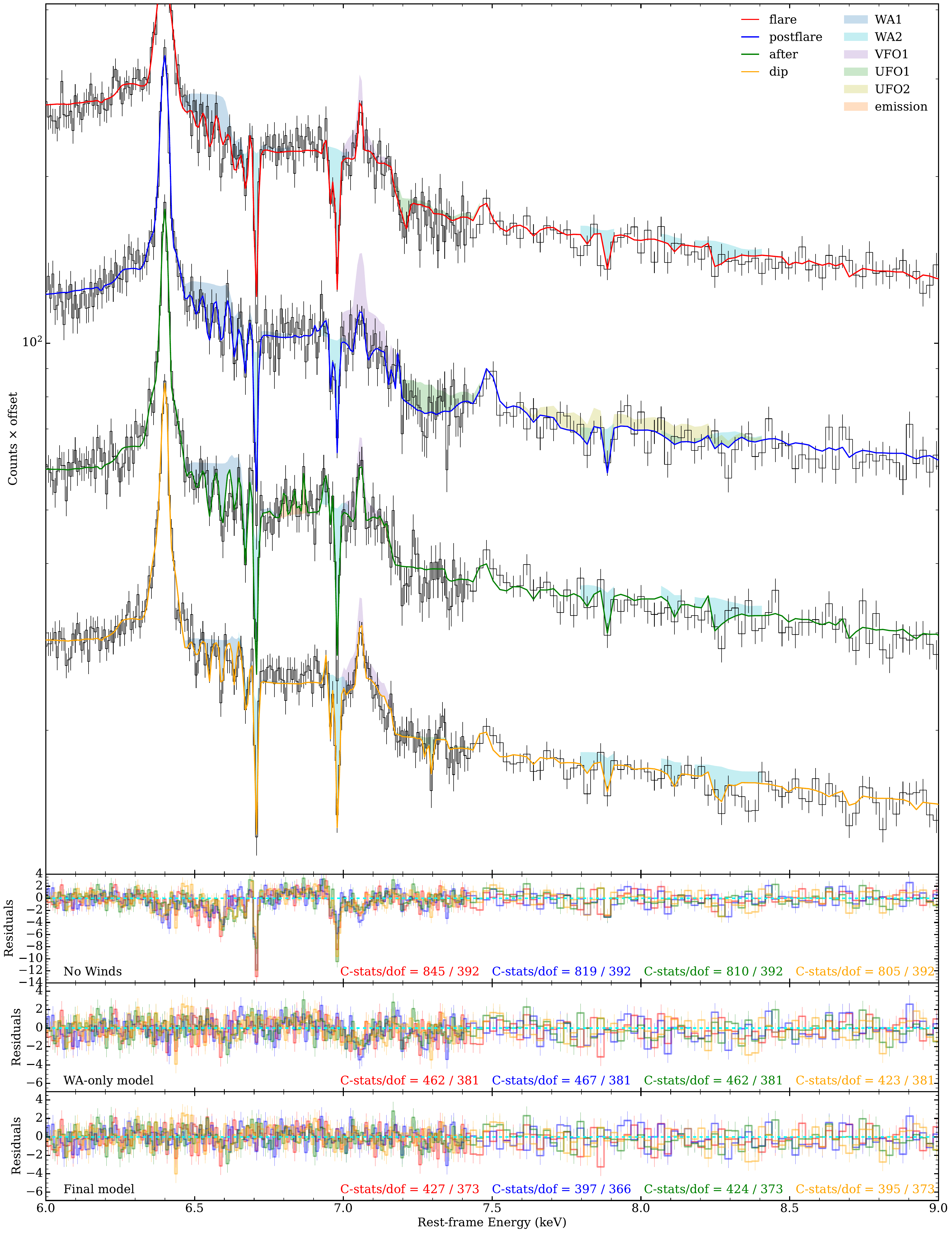}
    \caption{Top Panel: Spectra corresponding to time-selected phases (Flare, Postflare, After, and Hard-dip) with their best-fit models shown in matching colors (red, blue, green, and orange). Cyan dotted curves show the “No Winds” models obtained by removing all pion components and refitting the power-law normalization. Residuals Panels: Residuals for the No Winds model, the WA-only model (retaining only the warm absorber components and refitting their free parameters together with the power-law normalization), and the final model.}
    \label{fig:flares_fit}
\end{figure*}

\begin{figure*}
    \centering
    \includegraphics[width=0.95\linewidth]{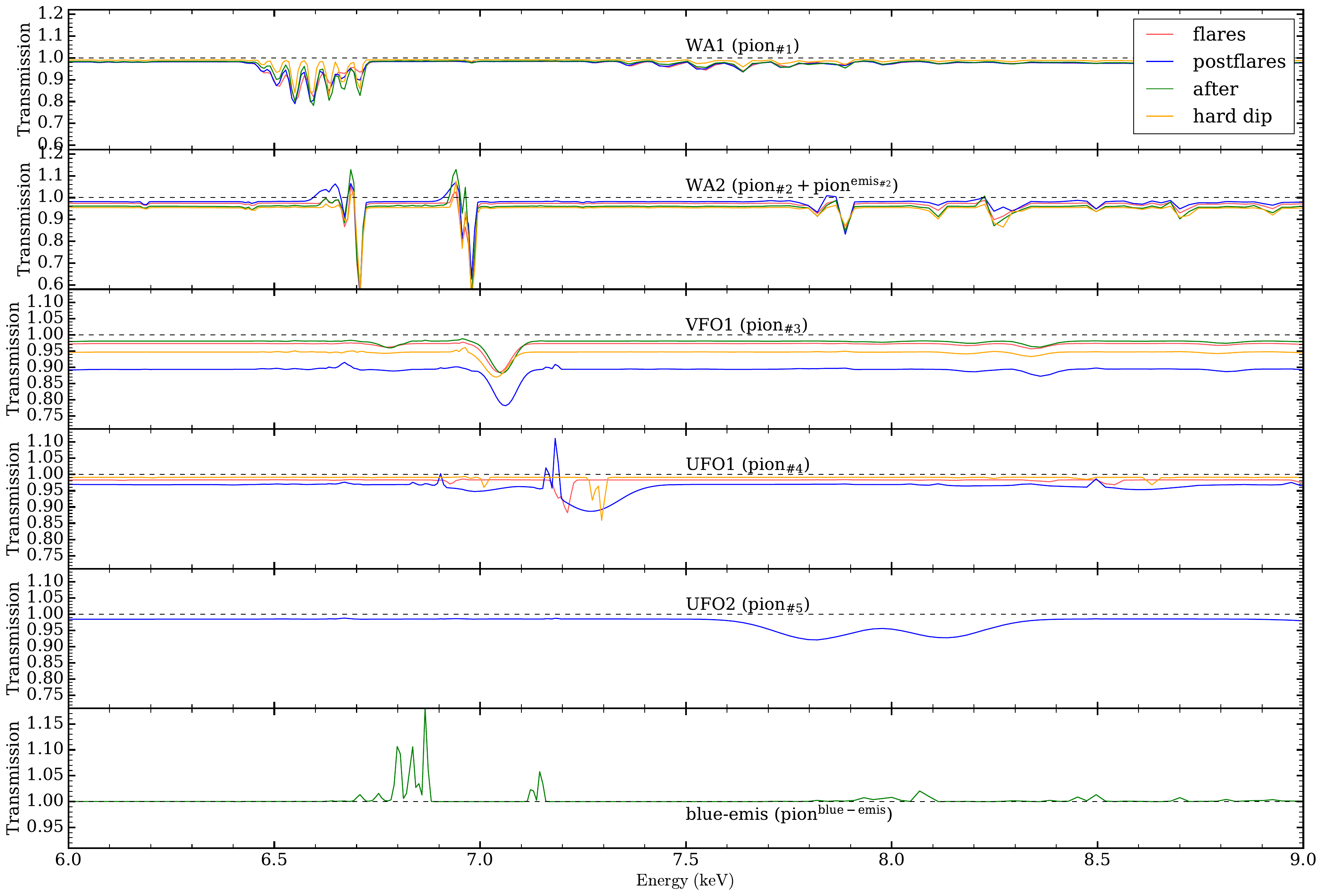}
    \caption{Transmission profiles of the individual ``pion" components for spectra of time-selected phases, computed by dividing the final model by the model with the specified component removed. Note, the marginally significant components ($1.5\sigma < \mathrm{D.S.} < 3\sigma$) include UFO1 in the flare spectrum and UFO1 in the hard-dip spectrum.}
    \label{fig:flares_trans}
\end{figure*}

\begin{figure*}
    \centering
    \includegraphics[width=1.0\linewidth]{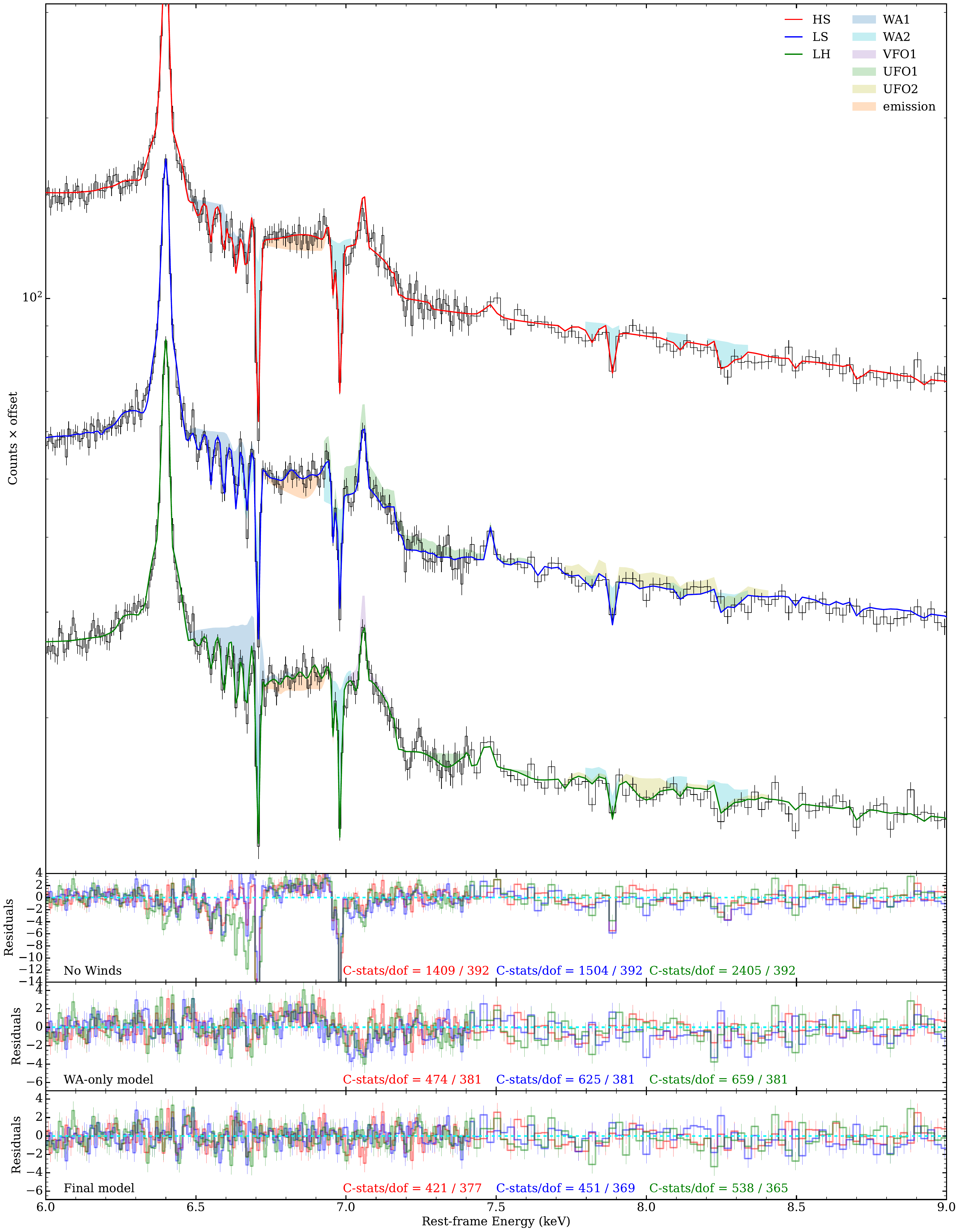}
    \caption{Same as Figure \ref{fig:flares_fit}, but for spectra grouped by global spectral regions (HS, LS, LH). The marginally significant component ($1.5\sigma < \mathrm{D.S.} < 3\sigma$) is UFO1 in the LH state.}
    \label{fig:regions_fit}
\end{figure*}

\begin{figure*}
    \centering
    \includegraphics[width=0.95\linewidth]{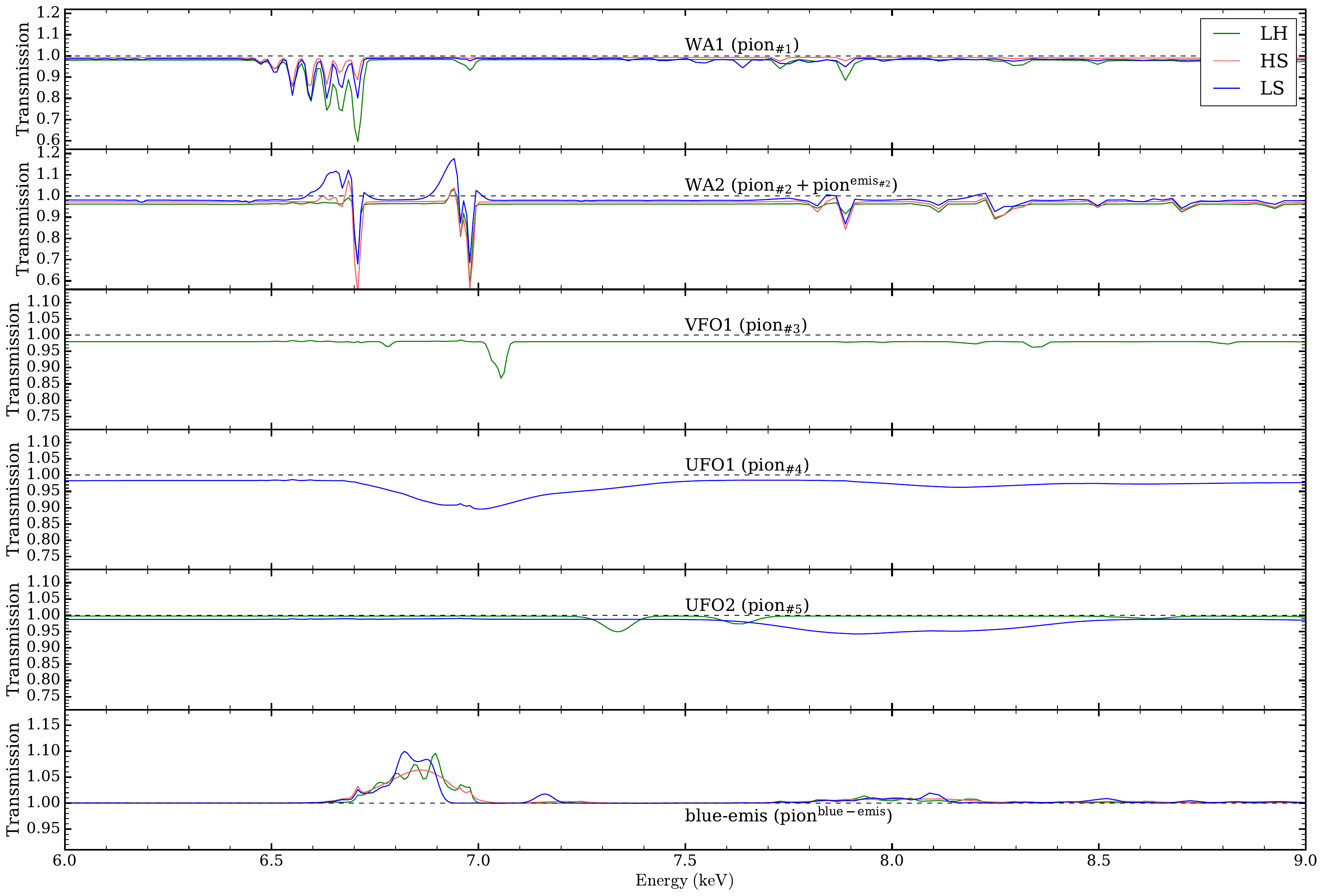}
    \caption{Transmission profiles of the individual ``pion" components for spectra of HID-selected regions.}
    \label{fig:regs_trans}
\end{figure*}

Figure \ref{fig:brd} presents the seven XRISM/Resolve spectral windows, each fitted independently over the 2.4–17.4 keV bandpass, with the best-fit models listed in Table \ref{table:parameters}. Figures \ref{fig:flares_fit}, \ref{fig:flares_trans}, \ref{fig:regions_fit}, and \ref{fig:regs_trans} show detailed views of the fits and the transmission profiles of the wind components in the 6.0–9.0 keV range. The reported Cash statistics and degrees of freedom (C-stat/dof) are given in the figures for the “No Winds” and “WA-only” models, to reflect the importance of fast winds in each spectral window.  These fit statistics were calculated by removing the relevant pion components and refitting the data, with the remaining wind parameters and power-law normalization allowed to vary freely.  The properties of the wind components in the final model are presented in the following section. The lower four panels of Figure \ref{fig:properties} show that the wind components vary with the selected spectra.

\subsection{Wind Components} \label{sec:prop}
We measured the significance of wind zones in each window by removing a given zone, refitting the spectrum, and evaluating the changes in the Cash statistic, AIC, and detection significance.  This procedure is more conservative than evaluating the significance of additional zones as a more complex model is assembled, because a larger accumulation of components are better able to model features than relatively few.  Here, we only record a wind zone within a window when the detection significance exceeds the $3\sigma$ threshold (D.S. $\geq 3\sigma$, defined in Appendix \ref{adx:DS})  The spectrum from every window requires two WA zones; these vary minimally.  The most complex spectrum requires five wind zones -- this is the spectrum of the ``Postflare'' window; it contains two WA zones, one VFO zone, and two UFO zones.  In contrast, the simplest spectra were obtained in the ``flare'' and ``HS'' windows -- both of these windows lack significant UFOs.  We describe the detailed observational properties of each wind component below.

\subsubsection{Warm Absorbers}
In every spectrum, the “zig-zag”-shaped absorption features between 6.45--6.65 keV are identified with transitions from Fe XX–XXIV (6.50, 6.54, 6.59, 6.63, and 6.66~keV). Stronger lines from Fe XXV–XXVI are simultaneously detected, including Fe XXV He$\alpha$ (6.70 keV) and He$\beta$ (7.88 keV), as well as Fe XXVI Ly$\alpha_{1,2}$ (6.95 and 6.97 keV) and Ly$\beta$ (8.25 keV). The lines from these charge states have modest blue-shifts, corresponding to outflow velocities of $-v \sim300~{\rm km~s^{-1}}$.  Two warm absorber zones are required to reproduce these slow absorption features (``$pion_{\#1}$'' and ``$pion_{\#2}$''). During the fits, significant re-emission in the red wings of the Fe XXV and Fe XXVI lines must also be included. We therefore add a broadened emission component ($vgau*pion^{emis\#2}$), in which the column density, ionization parameter, and turbulent velocity are linked to the corresponding absorption component. The emission covering factor, bulk velocity shift, and Gaussian broadening are left free.

The outer WA component (``$pion_{\#1}$'') is detected at high significance in all spectra (D.S. $> 7\sigma$ and $\Delta AIC < -60$ in all cases). It has column densities of $N_{\rm H} \sim (1.4$--$3.8)\times10^{22}~{\rm cm^{-2}}$ and ionization parameters of $\log\xi \sim 2.5$--$2.6$. The velocity remains stable across all event types ($- v_z \sim 200-310~{\rm km~s^{-1}}$), with modest turbulent velocities of $\sigma_v \sim 200$--$470~{\rm km~s^{-1}}$. No strong variability is observed in this component, suggesting a persistent, large-scale absorbing structure.

The inner WA component (``$pion_{\#2}$'') is also detected in all spectra with very high significance (D.S. $>8\sigma$ and $\Delta AIC<-70$ in all cases). It shows systematically higher ionization ($\log\xi \sim 2.9$--$3.3$) and higher column densities, $N_{\rm H} \sim (0.6$--$1.3)\times10^{23}~{\rm cm^{-2}}$. The velocity is consistent with the outer WA ($-v_z \sim 260-330~{\rm km~s^{-1}}$), but the turbulent velocity is lower ($\sigma_v \sim 130$--$170~{\rm km~s^{-1}}$). Compared to $pion_{\#1}$, this component shows some variability in both column density and ionization, with higher columns measured within the LH and hard-dip phases, suggestive of a closer connection to changes in the observed brightness and spectral shape.

Overall, the WAs are persistent across all states and phases, with relatively stable velocities but moderate variability in ionization and column density. This supports a picture in which the WA represents a long-lived, potentially ``failed'' wind seen close to apocenter, that responds weakly to short-timescale variability \citep{Miller_2026}.

\subsubsection{Very Fast Outflows}
VFOs (``$pion_{\#3}$'') are detected with strong detection significance (D.S. $\sim 3$--$5\sigma$ and $\Delta AIC<-8$) in a subset of spectra, but are absent ($<1\sigma$) in the HS and LS windows.  When these flows are detected, they have velocities of $-v_z \sim 2$--$4 \times10^3~{\rm km~s^{-1}}$, placing them between the WA and UFO regimes.  Most importantly, the column density of the VFOs in NGC~4151 is highest in the Postflare window ($N_{H}\sim3\times10^{23}~{\rm cm^{-2}}$), and lower in the After window ($N_{\rm H} \sim 5\times10^{22}~{\rm cm^{-2}}$).  In all cases, the ionization parameter of the observed VFO components is high, at $\log\xi \sim 3.5$--$3.8$.  However, whereas the VFO components require relatively high velocity broadening within the Postflare windows ($\sigma_v \sim 1000~{\rm km~s^{-1}}$), the broadening in the LH window is much lower ($\sigma_v \sim 300~{\rm km~s^{-1}}$). 

The rapid enhancement in the column density of the VFO components following flaring events suggests that they are closely linked to the inner accretion flow. The flare might trigger an additional physical response that enhances the launching of the gas. This physical response could be magnetic reconnection similar to NGC 3783 \citep{Vardi_2026}, propagation of magnetosonic waves, or ionization evolution of the gas. In this picture, the relevant magnetic loop or coronal disturbance whose characteristic causal scale set by the observed delay of 10 ks. A corresponding light-crossing scale of $\simeq 60r_g$ could be the most compact causal scale connecting the flare--wind response. At least qualitatively, this scenario is similar to some treatments of distributed, magnetically-dominated coronae (e.g., \citealt{hmg1994}, \citealt{tdm1998}, \citealt{merloni2001}), though alternative radial estimates are possible and discussed in Section \ref{sec:derivedprop}.

A second limiting possibility is the radial motion of the gas into and out of our line of sight. In this case, the relevant length scale is $v t_{flow}$. For the observed VFO velocities, $-v_z\sim2-4\times10^3~{\rm km~s^{-1}}\sim0.007-0.013c$, the distance traveled in 10 ks is only $\sim0.4-0.7~r_g$. This is much smaller than the $\sim6-10~r_g$ scale expected for a compact corona. Conversely, crossing a $\sim6-10~r_g$ X-ray source in $\sim10$ ks would require a transverse velocity of order $\sim0.1c$ (possible for a UFO), much larger than the observed line-of-sight VFO velocity. Thus, radial motion alone cannot easily explain the enhanced absorption. This difficulty may be alleviated if the true velocity is larger because of projection, or if transverse or Keplerian motion dominates the source crossing, or if this VFO represents gas that is lifted into view but not fully accelerated, or a ``failed'' wind.

If the VFO is at a radial distance of $\simeq 60r_g$ from the black hole, the central engine is not a point source, and orbital velocities will be imprinted within absorption lines, in a degree that is determined by the ratio of the central engine size and absorber size.  The projected local Keplerian velocity at this radius is $v \simeq 2\times 10^{4}~{\rm km}~{\rm s}^{-1}$ (assuming that the inner flow is observed a low inclination, $\theta \leq 30$~degrees), and the escape velocity is 40\% higher.  Yet, the VFO bulk velocities and velocity broadening are only about 10\% of these estimates.  If the central engine and clumps within the VFO both have a characteristic size of $r \simeq 6-10~GM/c^{2}$, velocity broadening close to the low observed values is just possible.  However, the contrast between the low observed bulk velocities and high local escape velocities is more difficult to reconcile.  Unless VFOs are relativistic surface flows along the plane of the disk -- a premise that is at odds with wind driving mechanisms -- they may also be ``failed'' winds that do not escape.

\subsubsection{Ultra Fast Outflows}
The UFO zones (``$pion_{\#4}$'' and ``$pion_{\#5}$'') trace the fastest and most highly variable wind components observed in NGC~4151.  The slower zone (``$pion_{\#4}$'') has velocities between $-v_{z} \simeq 1.0-2.7\times 10^{4}~{\rm km}~{\rm s}^{-1}$ (roughly 0.03--0.1c), while the faster zone (``$pion_{\#5}$'') has velocities of $-v_{z} \simeq 4.1-4.8\times 10^{4}~{\rm km}~{\rm s}^{-1}$ (roughly 0.14--0.16c).

The fast UFO zone (``$pion_{\#5}$'') is significantly detected only in Postflares, LS, and LH spectrum. It is not required in the Flare, where its detection significance is $<1\sigma$, and the listed value of $N_{\rm H}=1.5^{+0.7}_{-0.7}\times10^{22}~{\rm cm^{-2}}$ should therefore be interpreted as an upper-limit/sensitivity estimate rather than a measured column density. In contrast, $pion_{\#5}$ is significantly required in the initial Postflare window, where it reaches $N_{\rm H}=3.8^{+0.7}_{-0.6}\times10^{22}~{\rm cm^{-2}}$ with D.S. $=3.6\sigma$ and $\Delta{\rm AIC}=-11$. Thus, the main variability of $pion_{\#5}$ is not simply a modest change in column density among three detections, but rather the appearance of a statistically required absorber in the Postflare spectrum and its non-detection before and after.

Similar to the VFO zones, the fast UFO zone (``$pion_{\#5}$'') is absent (D.S $< 1\sigma$) or weak during the Flare window ($N_{H} = 1.5^{+0.7}_{-0.7}\times 10^{22}~{\rm cm}^{-2}$), strongest within the Postflare window ($N_{H} = 3.8^{+0.7}_{-0.6}\times 10^{22}~{\rm cm}^{-2}$), and absent in the After window. The slower UFO zone (``$pion_{\#4}$'') is also strongest in the Postflare window ($N_{H} = 8.6^{+1.1}_{-1.1}\times 10^{22}~{\rm cm}^{-2}$) and absent in the After window.

A corresponding Gaussian broadened emission component (``$vgau*pion^{emis\#4}$'') associated with $pion_{\#4}$ is only significantly required during the initial post-flare window, where the $pion_{\#4}$ UFO is also strongest. The emission exhibits a velocity of $v_z \sim -9000~{\rm km~s^{-1}}$, slower than the absorption component ($v_z \sim -1.3\times10^4~{\rm km~s^{-1}}$), and a covering factor of $\Omega \sim 0.10$, indicating a global extent of the outflow. The presence of this emission component suggests that we captured the strongest phase of the slow UFO. It not only intercepts the line of sight but also geometrically extends.

Here again, this indicates that the UFO zones, at least, are very sensitive to flaring. They could also be at a region as small as that inferred for the VFOs ($r\leq 60~GM/c^{2}$) if the 10ks interval is interpreted as the same causal limit, and may also be clumpy to explain the modest velocity broadening that is observed. 

The observed UFO velocities and a flow time of 10 ks imply traveled distances comparable to plausible compact coronal scales. The slower Postflare UFO, with $v_z\sim0.03-0.1c$, would move by $\sim15-50~r_g$ in 10 ks, while the faster UFO, with $v_z\sim0.14-0.16c$, would move by $\sim70-80~r_g$. Thus, unlike the VFOs, the UFO velocities are large enough that a vertical-crossing or flow-time interpretation is at least geometrically plausible.

However, whereas the bulk velocity shift of the VFOs is too low to escape to infinity (potentially making them failed winds), the fast UFO zone displays velocities that are comparable to local Keplerian velocities and escape speeds, likely signaling that this zone escapes to infinity and contributes significant feedback to the host galaxy. 

The measured columns and significance of the UFO components are also strong within the LS and LH windows, but the values are consistent with non-detections in the HS window and hard dips. This may signal that a UFO is produced not only in response to flares, but also in a process that occurs when the ionizing flux is below a given threshold. The data may signal that over-ionization may reduce the production of UFOs.

Given the broadly similar behaviors and dependencies of the VFOs and UFOs, it is possible that they are produced via the same process (driven by the same mechanisms).  The UFO components may represent parts of an inner disk wind that escape to infinity and alter the host galaxy, while the VFOs may be parts of the same flow that were not launched or accelerated to the local escape speed.  Given that magnetic processes appear to play an important role in launching UFOs (see, e.g., \citealt{fukumura2018}, \citealt{Xiang_2025}, \citealt{Gu_2025}), the difference may be down to the magnetic field strengths and the mass of the clumps that may experience magnetic force.

\begin{figure*}
    \centering
    \includegraphics[width=1\linewidth]{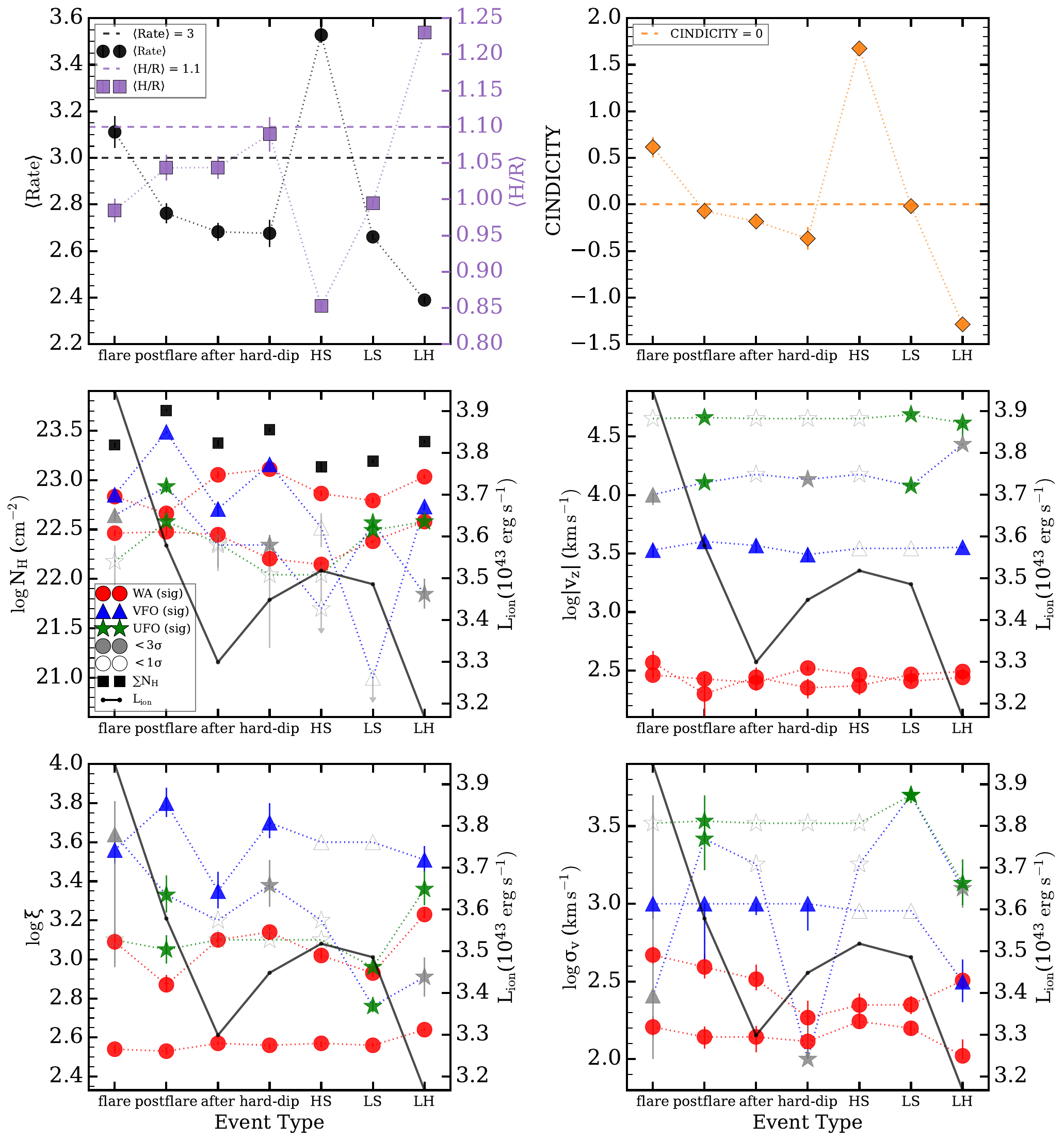}
    \caption{The spectral and wind properties as a function of event type. The top left panel shows the mean count rate and hardness ratio. The top right panel shows the CINDICITY. The lower four panels show the wind column density ($\log{N_H}$), ionization parameter ($\log{\xi}$), line-of-sight outfow velocity ($\log{|v_z|}$), and turbulent velocity ($\log{\sigma_v}$). Circles, triangles, and stars represent WA, VFO, and UFO components, respectively, with dotted lines connecting the same component number across event types, as indicated in Table \ref{table:parameters}. Filled colored markers indicate significant detections (D.S > 3$\sigma$), filled gray markers indicate marginal detections (1$\sigma$ < D.S < 3$\sigma$), and open gray markers indicate weak/no detections (D.S < 1$\sigma$). Black squares in the second panel mark the summed column density. The black lines in the lower four panels show the total ionizing luminosity in each event type.}
    \label{fig:properties}
\end{figure*}

\begin{figure*}
    \centering
    \includegraphics[width=1.0\linewidth]{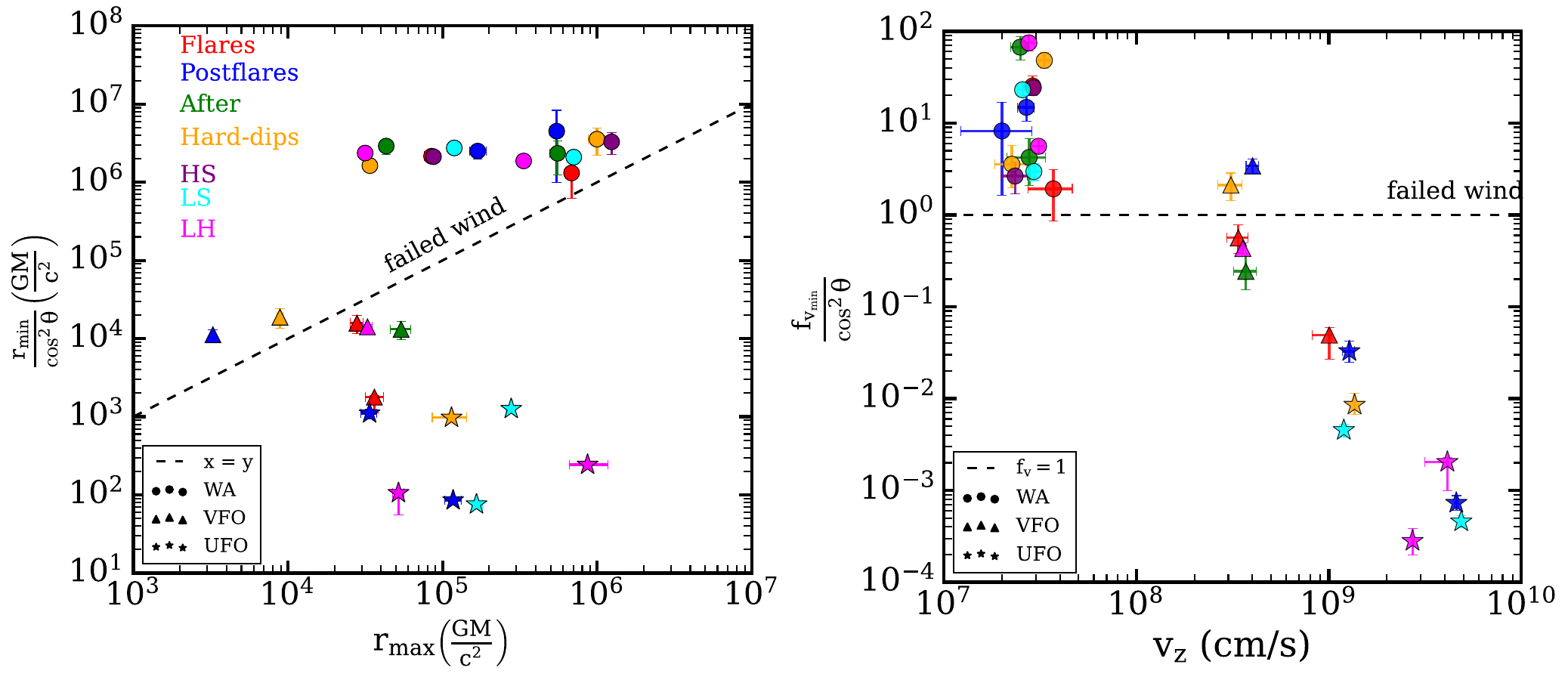}
    \caption{Left panel: comparison of the escape-based lower limit to the launching radius, $r_{\rm min}/\cos^2\theta$, and the ionization-based upper limit, $r_{\rm max}$, for each outflow component, in units of $GM/c^2$. The dashed line marks equality. Circles, triangles, and stars denote WAs, VFOs, and UFOs, respectively, while colors indicate different event types and HID-selected spectra. Components above the equality line formally require $f_v>1$. Right panel: minimum volume filling factor, $f_{v,{\rm min}}/\cos^2\theta$, as a function of velocity. The dashed line marks $f_v=1$. Faster outflows require smaller filling factors, with UFOs occupying the lowest values, VFOs intermediate values, and WAs clustering near or above unity.}
    \label{fig:radius}
\end{figure*}

\begin{figure}
    \centering
    \includegraphics[width=1.0\linewidth]{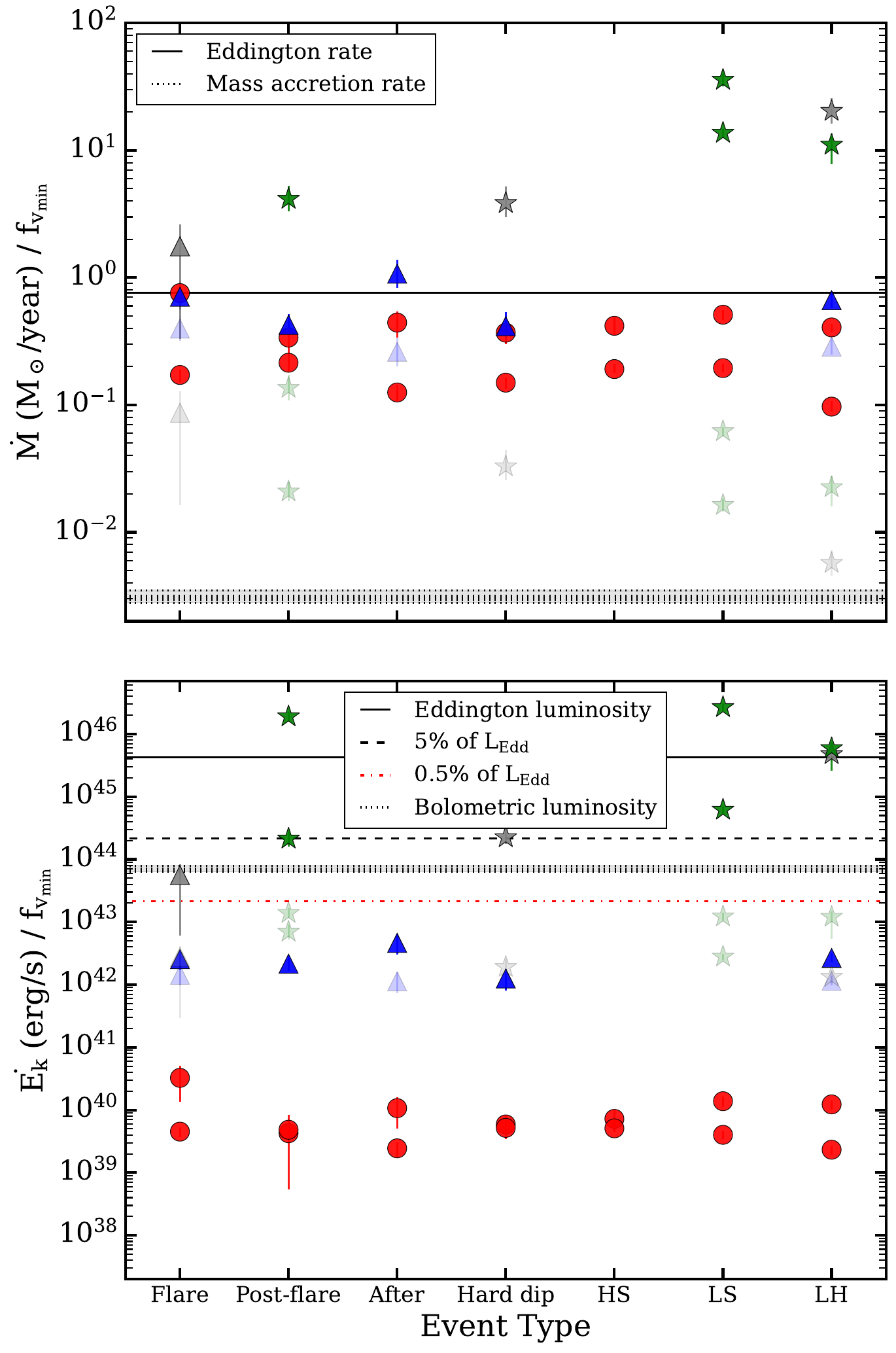}
    \caption{Upper panel: mass outflow rate, $\dot{M}/f_{v,{\rm min}}$, as a function of event type. Lower panel: kinetic power, $\dot{E}_{\rm k}/f_{v,{\rm min}}$, as a function of event type. Circles, triangles, and stars denote WAs, VFOs, and UFOs, respectively. Filled markers show the nominal values assuming $f_v=1$, while the lighter markers show the values corrected by the minimum volume filling factor (only when their $f_{v,{\rm min}<1}$). The grey markers show the components with $1\sigma <$ D.S < $3\sigma$. The solid horizontal lines mark the Eddington accretion rate (upper panel) and Eddington luminosity (lower panel). The dashed and red dotted lines in the lower panel mark 5\% and 0.5\% of $L_{\rm Edd}$, and the dotted band marks the bolometric luminosity range. WAs remain energetically weak, while VFOs and especially UFOs dominate the outflow energetics and can approach or exceed the nominal feedback threshold.}
    \label{fig:energetics}
\end{figure}

\subsubsection{Constant Column Test for Wind Variability}
The individual D.S. and AIC tests for each wind component have established whether one wind zone is statistically required in a given spectrum. To further test whether the same component can be present with a fixed column density across all seven spectra, we performed an additional set of fits for each of the five absorption components. For each tested component, we fixed its column density to the mean of the values measured in the spectra when it's significantly detected (D.S > $3 \sigma$). We then refit all seven spectra, allowing the remaining three wind parameters to be free as well as the powerlaw normalization. The results are summarized in Table \ref{tab:constant_NH_test}. 

All five wind components disfavor the constant column model with positive $\Delta AIC $. The two persistent WA zones, $pion_{\#1}$ and $pion_{\#2}$, shows high positive $\Delta AIC$. This is expected because both WA components are highly significant in every spectrum with well constrained column densities. Even a modest forced change away from the best-fit column leaves significant residuals. This is also suggested from the best-fit final model, with maximum differences of $\sim 17\sigma$ for $pion_{\#1}$ between HS and LH, and $\sim 8\sigma$ for $pion_{\#2}$ between Postflare and Hard-Dip, the WA zones show highly significant column changes. Similarly for the VFO component, $pion_{\#3}$, and the slower UFO components, $pion_{\#4}$, the fixed column model is not prefered. The VFO component $pion_{\#3}$ shows the measured change of $\sim 21\sigma$ between Postflare and After, while the slower UFO component $pion_{\#4}$ varies at $\sim7\sigma$ between LS and LH.

For the fastest UFO components, $pion_{\#5}$, the fixed column model gives a slightly lower Cash statistic, $\Delta C=-19$. This does not imply that a constant-column absorber is physically preferred. Rather,  $pion_{\#5}$ is not detected in four of the seven spectra, with D.S. $< 1 \sigma$. Adding a fixed-column component can slightly decrease the Cash statistic, even though the component is not independently detected. Once the additional degrees of freedom are penalized through the AIC, the fixed column model is still not preferred.  We therefore interpret $pion_{\#5}$ as a transient component whose detectability changes across spectral windows.

\begin{table*}
\centering
\caption{Constant-column variability test for the five absorption components.}
\label{tab:constant_NH_test}
\begin{tabular}{lcccccc}
\hline
\hline
Component 
& $\langle N_{\rm H}\rangle_{\rm sig}$
& $C_{\rm final}/\nu_{\rm final}$
& $C_{\rm fixed}/\nu_{\rm fixed}$
& $\Delta C/\Delta\nu$
& $\Delta{\rm AIC}$ \\
& $(10^{22}~{\rm cm^{-2}})$
& 
& 
& 
& \\
\hline
$pion_{\#1}$ & 2.56  & 3051/2596 & 3172/2603 & 121/7 & 105 \\
$pion_{\#2}$ & 8.56  & 3051/2596 & 3097/2603 & 46/7 & 30 \\
$pion_{\#3}$ & 12.5 & 3051/2596 & 3101/2595 & 50/-1 & 52 \\
$pion_{\#4}$ & 6.15  & 3051/2596 & 3083/2595 & 32/-1 & 34 \\
$pion_{\#5}$ & 3.57  & 3051/2596 & 3032/2587 & -19/-9 & 2 \\
\hline
\end{tabular}
\tablecomments{
The significant spectra are those with D.S. $>3\sigma$ in
Table~\ref{table:parameters}. We define
$\Delta C=C_{\rm fixed}-C_{\rm final}$ and
$\Delta{\rm AIC}={\rm AIC}_{\rm fixed}-{\rm AIC}_{\rm final}$, where $C_{\rm final}$ is the summed Cash statistic of the final adopted model over all seven spectra. Positive
values indicate that the fixed-column model is disfavored.
}
\end{table*}

\subsection{Intrinsic Obscuration} \label{subsec:hot}
%hot component
In addition to the ionized wind components, all seven windows require a cool partially covering absorber, as per \cite{Miller_2026}. The fitted properties of the ``hot'' component are summarized in Table~\ref{table:parameters}. 
In paper~I, the partially covering absorber was assumed to be neutral, and distant from the central engine.  The fits made to the seven windows with the revised model find best-fit temperatures of $T \simeq 3$--$5$ eV, signaling mild ionization.  This is consistent with the slightly shifted Fe K edge discussed in \cite{Miller_2026}.

The covering factor of this absorber remains high and nearly constant in all windows, with $f_{\rm cov}\simeq 0.85$--0.88.  This implies that the basic geometry of this component is relatively constant, and that it consistently covers most of the central engine.  The column density of this internal absorber does not vary much across the time windows, remaining within $N_{\rm H}\simeq (1.7$--$1.8)\times10^{23}~{\rm cm^{-2}}$ in the Flares, Postflares, Afters, and Hard-dips windows.  In contrast, a clearer trend emerges in the HID-selected windows: the column density is lowest in the HS state, $N_{\rm H}=1.29^{+0.02}_{-0.09}\times10^{23}~{\rm cm^{-2}}$, increases in the LS state to $N_{\rm H} = 1.78^{+0.20}_{-0.15}\times10^{23}~{\rm cm^{-2}}$, and is highest in the LH state at $N_{\rm H} = 2.01^{+0.05}_{-0.13}\times10^{23}~{\rm cm^{-2}}$.  This suggests that the intrinsic obscuration is more strongly linked to the global state of NGC~4151, than to the short-timescale local variability phases. The ``softer when brighter'' behavior may therefore be at least partly driven by absorption variability where increased obscuration can suppress the softer part of the Resolve band and make the source appear harder.

\begin{figure*}
    \centering
    \includegraphics[width=1.0\linewidth]{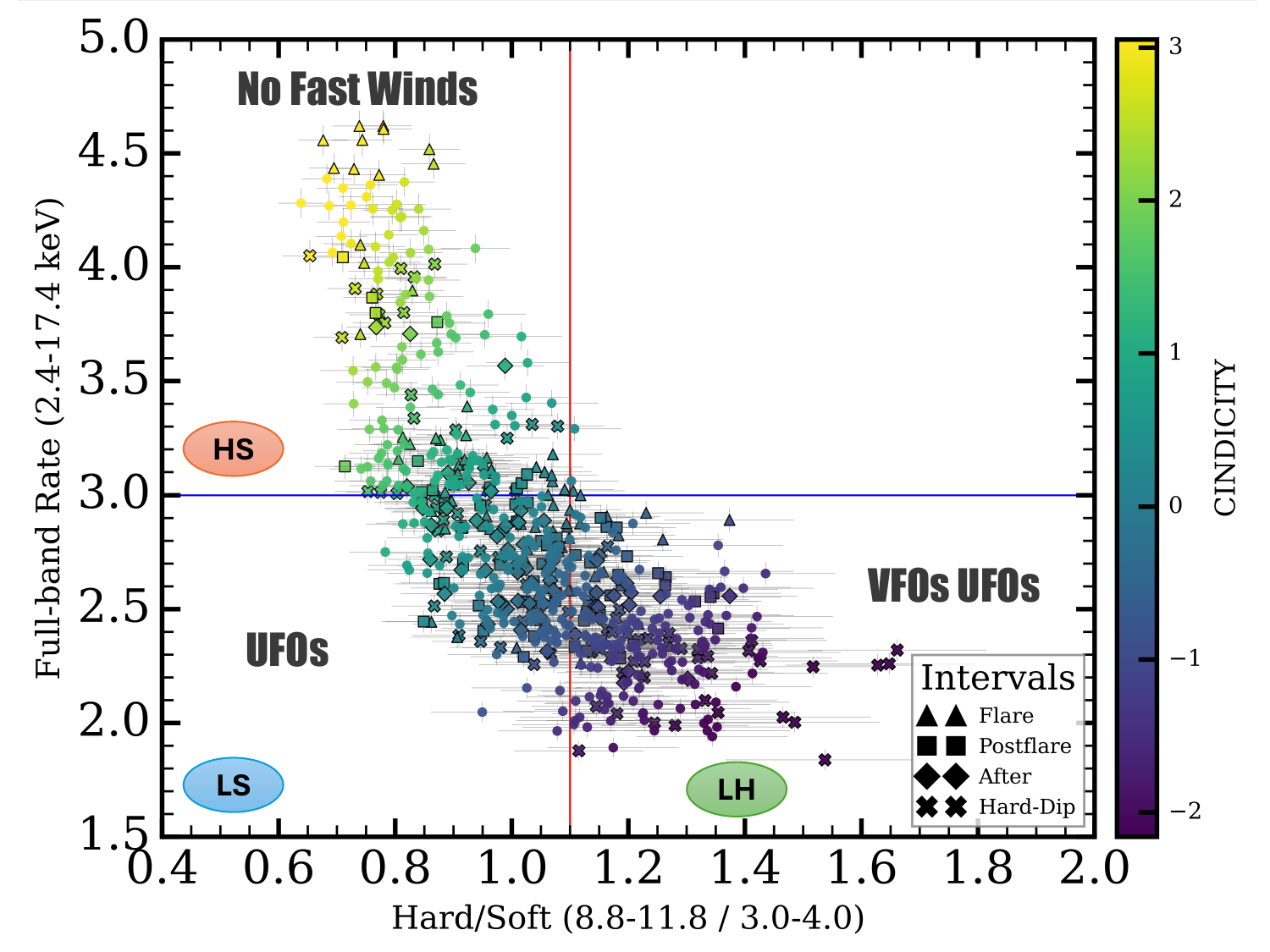}
    \caption{Hardness--intensity diagram for all time bins, colored by CINDICITY. The horizontal and vertical lines mark the boundaries used to define the global spectral states: High--Soft (HS), Low--Soft (LS), and Low--Hard (LH). Different symbols denote the time-selected intervals (Flare, Postflare, After, and Hard-dip). The labels indicate the regions in which no fast winds, only UFOs, or both VFOs and UFOs are preferentially detected.}
    \label{fig:HID_cind}
\end{figure*}

\begin{figure}
    \centering
    \includegraphics[width=1.0\linewidth]{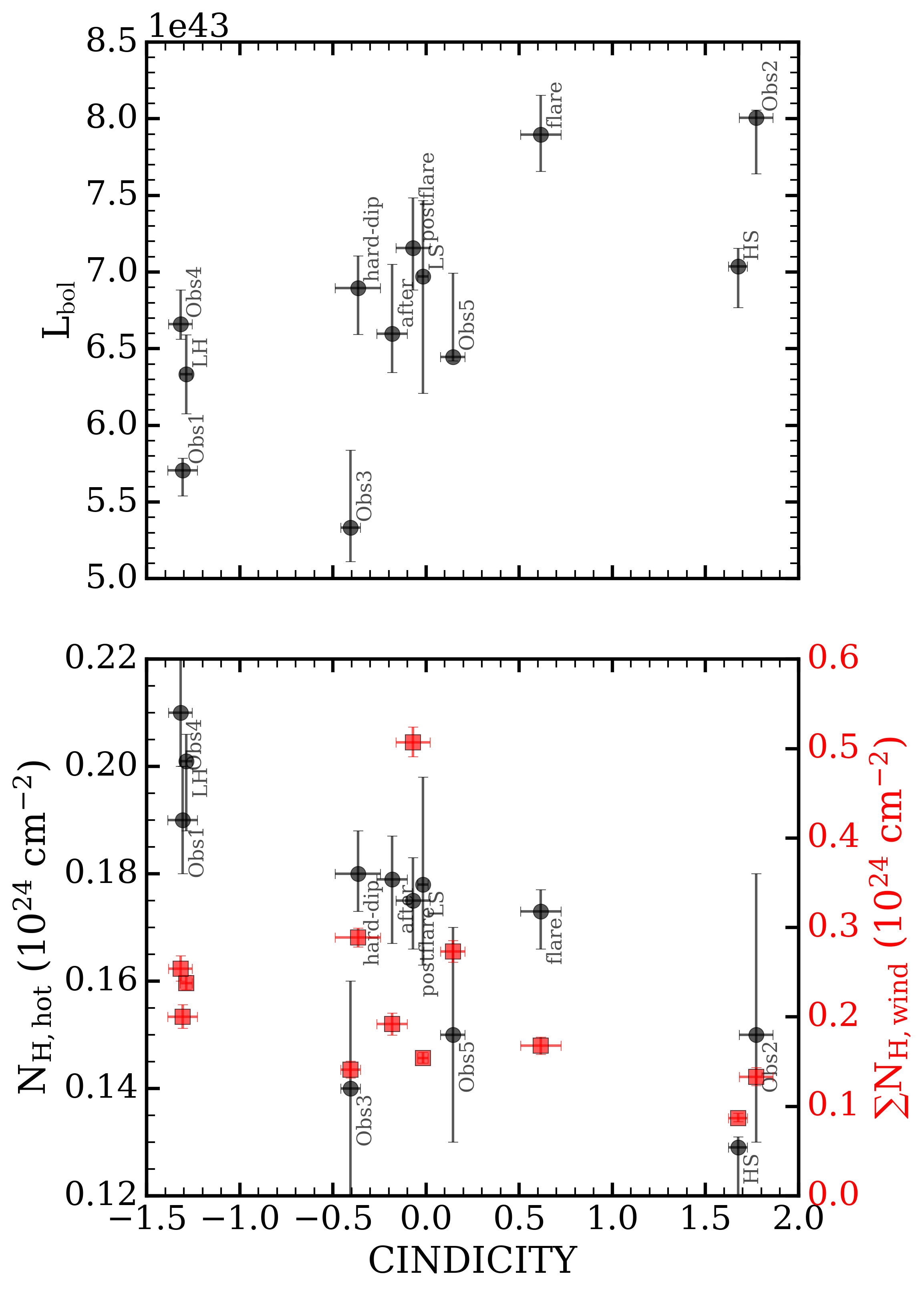}
    \caption{Top panel: bolometric luminosity, $L_{\rm bol}$, as a function of CINDICITY. Bottom panel: intrinsic obscurer column density, $N_{\rm H,hot}$ (black circles; left axis), and total wind column density, $\Sigma N_{\rm H,wind}$, for significant (D.S > $3 \sigma$) wind components (red squares; right axis), as a function of CINDICITY. The points include the seven time-selected and HID-selected spectra from this work together with the five PV observations from Paper~I. Dimmer, harder spectra correspond to smaller CINDICITY and stronger intrinsic obscuration, while the total wind column shows a weaker trend.}
    \label{fig:nH_cind}
\end{figure}

\subsection{Launching Radius and Energetics} \label{sec:derivedprop}
Following the same approach as in Paper~I, we use the best-fit ionization parameter, column density, velocity, and the ``$lixi$" output ($L_{\rm{ion}}/\xi$) of each absorption component to estimate characteristic radius limits, minimum volume filling factors, mass outflow rates, and kinetic powers. We do not fit for a parameter “lixi” in each component. It is internally calculated by the model, allowing us to estimate the ionizing luminosity incident on the inner edge of each absorbing layer. Multiplying by their best fit ionization parameters, the inferred ionizing luminosities are consistently within $2.9 - 3.7 \times 10^{43}~\mathrm{erg~s^{-1}}$. 

The ionization parameter is defined as
\begin{equation}
    \xi = \frac{L_{\rm ion}}{n r^2},
\end{equation}
where $L_{\rm ion}$ is the ionizing luminosity in the 13.6 eV--13.6 keV band, $n$ is the density of a gas clump, and $r$ is the distance from the black hole. If the radial thickness of an absorber does not exceed its distance from the source, $\Delta r \le r$, then with $N_{\rm H} = f_v  n \Delta r \le f_v n r$, the ionization condition gives an upper limit to the launching radius,
\begin{equation}
    r_{\rm max} = \frac{f_v L_{\rm ion}}{\xi N_{\rm H}} \equiv f_v r_1 ,
\end{equation}
where $f_v$ is the volume filling factor along the flow. A lower limit can be estimated by requiring that the observed outflow velocity not fall below the local escape velocity,
\begin{equation}
    r_{\rm min} = \frac{2GM}{v_{\rm out}^2}
    = \frac{2GM \cos^2\theta}{v_z^2}
    \equiv r_2 \cos^2\theta ,
\end{equation}
where $v_z$ is the observed line-of-sight velocity, $v_{\rm out}\cos\theta = v_z$, and $\theta$ is the inclination of the flow relative to the line of sight. Requiring $r_{\rm max}\ge r_{\rm min}$ gives the minimum volume filling factor,
\begin{equation}
    f_{v,{\rm min}} = \frac{r_2}{r_1}\cos^2\theta .
\end{equation}

Figure~\ref{fig:radius} summarizes these constraints for all time-selected and HID-selected outflow components. The left panel compares $r_{\rm min}/\cos^2\theta$ and $r_{\rm max}$ in units of $GM/c^2$. The WAs lie mostly above unity, implying formally $f_v>1$ if they are assumed to be observed at their launching radii. As in Paper~I, this suggests that the WAs are either seen at large $\theta$, observed far from their launching points, and/or correspond to failed winds. By contrast, the VFOs and UFOs generally satisfy $r_{\rm min}<r_{\rm max}$ and occupy progressively smaller radii. The right panel shows $f_{v,{\rm min}}/\cos^2\theta$ as a function of velocity. A clear trend is present: faster outflows require smaller minimum filling factors. The UFOs occupy the lowest values, implying highly clumpy flows, while the VFOs are intermediate and the WAs cluster near or above unity. For the points with $f_v>1$, the minimum filling factor loses its physical meaning, so we do not apply this correction when calculating the energetics below.

The characteristic radius limits above are very broad because they rely on simply geometric and escape-speed arguments. We next considered more scenarios to obtain a tighter limit considering the observed 10ks causal timescale.

First is to rule out the possible contributions from recombination timescales. We estimated recombination timescales directly from the SPEX calculation, which gives the recombination timescale per ion following the definition of \cite{Bottorff_2000}. For the Postflare VFO and UFO components, the assumed model density is $n_{\rm H}=10^9~{\rm cm^{-3}}$, with $n_e\simeq1.2\times10^9~{\rm cm^{-3}}$. At this density, the relevant Fe recombination timescales are only $\sim0.04$ ks for the VFO, $\sim0.18$ ks for the slower UFO, and $\sim0.19$ ks for the faster UFO.  These values are already more than an order of magnitude shorter than the adopted 10 ks Postflare window. Thus, the observed $\sim10$ ks Postflare behavior is unlikely to be set by the recombination time of the gas. Instead, the absorbing gas should remain close to photoionization equilibrium as the continuum varies. 

A more plausible interpretation for the fast Postflare variability is that the $<10$ ks interval reflects a causal timescale involving magnetic/coronal activity and mechanical evolution of the gas. For reference, a 10~ks light-crossing time corresponds to $60~r_g$ for NGC~4151. If the Postflare VFOs and UFOs are located on this compact limiting scale, the ionization parameter implies densities of $n = L_{ion}/(\xi r^2) = 5.8 \times 10^{10} \rm{cm^{-3}}$ and $f_v > N_H/nr =  0.018$ for the Postflare VFO, $n = 1.8 \times 10^{11} \rm{cm^{-3}}$ and $f_v > 1.6 \times 10^{-3}$ for the slower UFO, and $n = 3.4 \times 10^{11} \rm{cm^{-3}}$ and $f_v > 3.8 \times 10^{-4}$ for the faster UFO.  These densities are broadly consistent with the base densities in MHD disk-wind calculations \citep{Fukumura_2015}. Using these densities and scaling them as $t_{\rm rec}\propto n_{\rm H}^{-1}$, the recombination times are $\lesssim0.8$ s for the VFO, $\lesssim1$ s for the slower UFO, and $\lesssim0.6$ s for the faster UFO. Even if the effective recombination coefficient were two orders of magnitude smaller, the recombination times would remain $\lesssim$100 s, still well below the 10 ks variability time. Thus, in the compact limiting scale, the gas can remain in photoionization equilibrium on 10 ks timescales. 

The inferred volume-filling factors from the compact limiting scale are generally smaller than the $f_{v,min}$ for the two UFOs, implying that if the $<10$ ks variability is causal, the UFOs can be more clumpy. For the Postflare VFO, the condition $f_{v,\min}>1$, shown in Figure \ref{fig:radius}, disfavors a freely escaping wind launched at the inferred radius. Taken together with the compact causal limit implied by the $\sim10$ ks flare--wind response, this supports a picture in which the VFO traces dense inner-wind clumps that are only partially accelerated and may ultimately fail to escape.

An alternative order-of-magnitude radial estimate motivated by the size of the X-ray source. If the absorber thickness is comparable to the compact corona $\Delta R\sim 10r_g$, this gives an averaged gas density of $n\sim N_H/\Delta r \sim 10^9$ for a representative column of $N_{\rm H}\sim10^{23}~{\rm cm^{-2}}$. Combining this density with the measured $L_{ion}/\xi$ values through $r \sim (L_{\rm ion}\xi n)^{1/2}$ places the Postflare VFO/UFO component at characteristic radii of $10^2 - 10^3 r_g$, about an order of magnitude further than the compact limiting scale.

The mass outflow rate is calculated as
\begin{equation}
    \dot{M}_{\rm out}
    = 4\pi \mu m_p f_{\rm cov} f_v \frac{L_{\rm ion}}{\xi} v_{\rm out},
\end{equation}
where $\mu=1.23$ is the mean atomic weight, $m_p$ is the proton mass, and we adopt $f_{\rm cov}=0.5$ for the absorbers. Assuming the gas has reached its terminal velocity, the kinetic power is
\begin{equation}
    \dot{E}_{\rm k} = \frac{1}{2}\dot{M}_{\rm out} v_{\rm out}^2 .
\end{equation}
We further estimate the bolometric luminosity as $L_{\rm bol}=2L_{\rm ion}$ and calculate the mass accretion rate as
\begin{equation}
    \dot{M}_{\rm acc} = \frac{L_{\rm bol}}{\eta c^2},
\end{equation}
adopting $\eta=0.4$ for a rapidly spinning black hole \citep{Keck_2015}. The Eddington accretion rate is defined as $\dot{M}_{\rm Edd}=L_{\rm Edd}/\eta c^2$.

Figure~\ref{fig:energetics} shows the resulting $\dot{M}_{\rm out}$ and $\dot{E}_{\rm k}$ as a function of window type. The nominal values (with $f_v=1$) show stratification: WAs carry the lowest mass fluxes and kinetic powers, VFOs are intermediate, and UFOs are the most energetic. In particular, the UFOs in the Postflare and LS reach the largest nominal mass outflow rates and kinetic powers when the fast wind is strongest. WAs stay far below the canonical feedback threshold of $0.5\%\,L_{\rm Edd}$, whereas the VFOs and especially the UFOs approach or exceed the level in their nominal estimates. Thus, as in Paper~I, the slow absorbers are unlikely to drive strong galaxy-scale feedback, while the fast outflows remain the most plausible channel for significant kinetic coupling to the surrounding environment. Applying the minimum filling factors lowers the inferred $\dot{M}_{\rm out}$ and $\dot{E}_{\rm k}$ substantially for the VFOs and UFOs, but they remain the dominant contributors for the kinetic power. Unlike the UFOs in Paper~I, however, none of the UFOs here remain above the $0.5\%\, L_{\rm Edd}$ threshold under the most conservative estimates, while reaching marginal values of order \(\sim 0.3\%\,L_{\rm Edd}\). The feedback threshold is therefore only reached occasionally in NGC 4151, rather than persistently.

\begin{figure}
    \centering
    \includegraphics[width=1\linewidth]{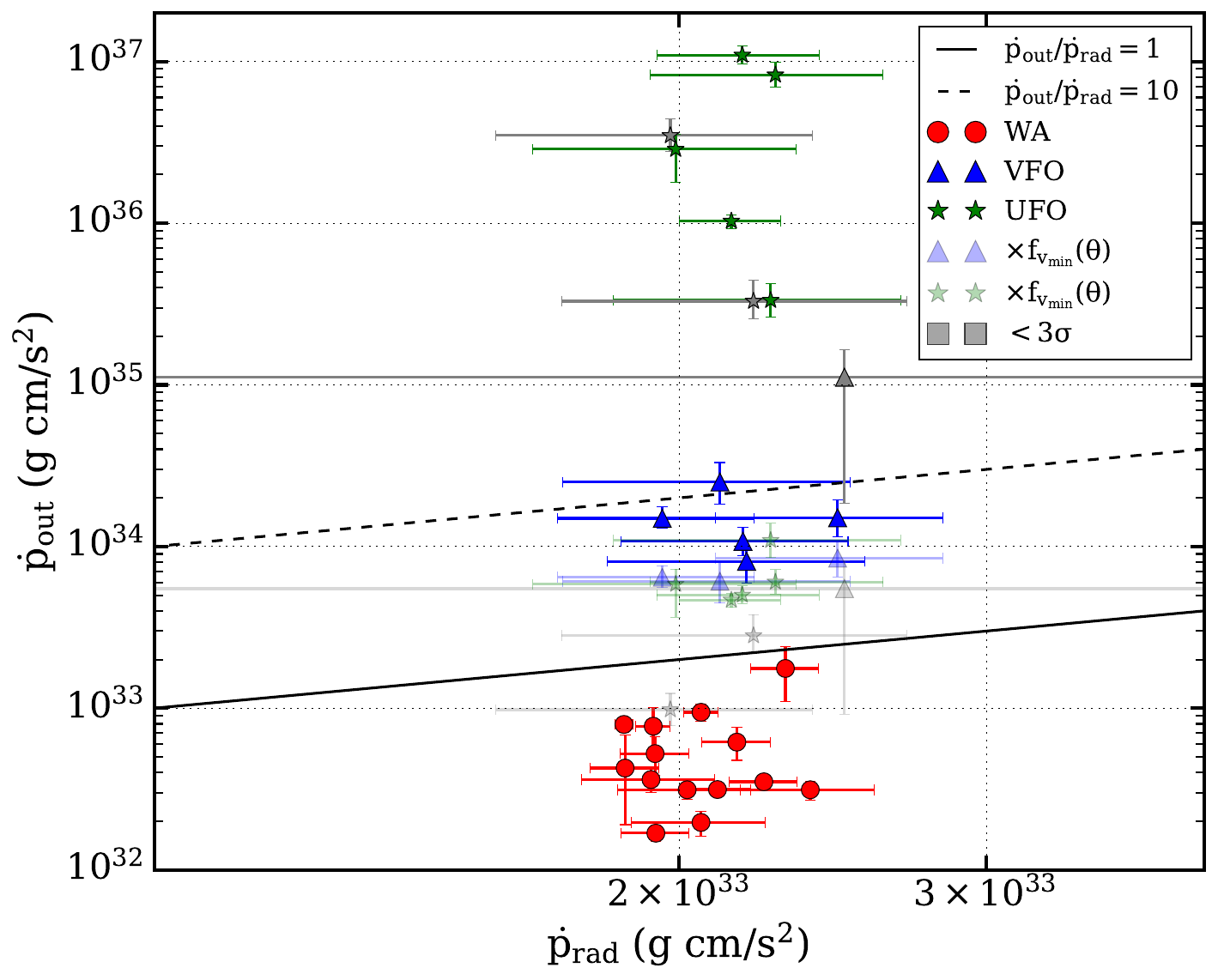}
    \caption{Outflow momentum rate versus radiation momentum flux, for all absorption components. Circles, triangles, and stars denote WAs, VFOs, and UFOs, respectively. Filled markers show the nominal values assuming $f_v=1$, while the lighter markers show the values after applying the minimum filling-factor correction, $f_{v,\min}(\theta)$, where physically meaningful. Gray markers indicate components with detection significance $<3\sigma$. The solid and dashed lines mark $\dot{p}_{\rm out}/\dot{p}_{\rm rad}=1$ and 10, respectively. WAs cluster below or near the momentum-conserving limit, VFOs occupy an intermediate regime, and UFOs are the most momentum-loaded components, often exceeding the radiative momentum supply in their nominal estimates.}
    \label{fig:momentum}
\end{figure}

Figure~\ref{fig:momentum} compares the outflow momentum rate, $\dot{p}_{\rm out}=\dot{M}_{\rm out}v_{\rm out}$, with the radiation momentum flux, $\dot{p}_{\rm rad}=L_{\rm bol}/c$, for all absorption components. As in Paper~I, the WAs lie below or near the $\dot{p}_{\rm out}/\dot{p}_{\rm rad}=1$ line, indicating that their momentum budget is broadly consistent with radiative driving. The VFOs occupy an intermediate regime, mostly between $\dot{p}_{\rm out}/\dot{p}_{\rm rad}=1$ and 10, while the UFOs generally exceed $\dot{p}_{\rm out}/\dot{p}_{\rm rad}=10$ in their nominal estimates, implying that radiation pressure alone is unlikely to explain the most extreme fast outflows. Applying the minimum filling-factor correction shifts the VFOs and UFOs downward, but the UFOs remain the most momentum-loaded components. Thus, consistent with the energetic trends above, the fastest winds are the most dynamically important, even if their strongest feedback signatures are only occasional.

\subsection{Global Trends among significant wind components} \label{sec:globaltrends}

The time-selected and HID-selected spectra suggest that both the continuum state and the wind properties evolve together, but the full set of trends is difficult to assess using the hardness--intensity diagram alone. To facilitate a more uniform comparison and understand \textit{how} the winds are launched, we introduce the scalar quantity \emph{CINDICITY}, defined in Appendix \ref{adx:CIND}, which combines count rate and hardness ratio into a single coordinate along the main direction of source evolution in the HID. The top two panels of Figure \ref{fig:properties} show the mean count rate, hardness ratio, and the CINDICITY varies with selected spectra. Previous Swift/XRT monitoring also showed that NGC~4151 exhibits strong hardness variability, including ``harder when brighter'' behavior when hardness is defined over the softer 0.3--10 keV XRT band \citep{Peretz_2018}. However, the XRISM/Resolve hardness ratio used here probes a harder bandpass, so the two trends are not directly equivalent.

In our convention, larger CINDICITY corresponds to brighter and softer conditions, while smaller CINDICITY corresponds to fainter and harder conditions. As shown in Figure~\ref{fig:HID_cind}, the HS spectrum lies at the bright-soft end of the sequence and is associated with no fast winds, whereas the LS and LH spectra lie at lower CINDICITY and preferentially host VFO and UFO components. The lower panel of Figure~\ref{fig:nH_cind} further shows that intrinsic obscuration $N_{\rm H,hot}$ is anti-correlated with CINDICITY, while the total wind column density shows a weaker trend. The ``softer when brighter'' behavior captured by CINDICITY may therefore be dominated by short-timescale soft flaring in the Resolve band, while the harder-when-dimmer/LH states may be more strongly shaped by obscuration-driven variability. The absence of fast winds in the HS spectrum may also be associated with the short-time flare state itself, since a larger portion of the flare events are in the HS region.

\begin{figure*}
    \centering
    \includegraphics[width=1.0\linewidth]{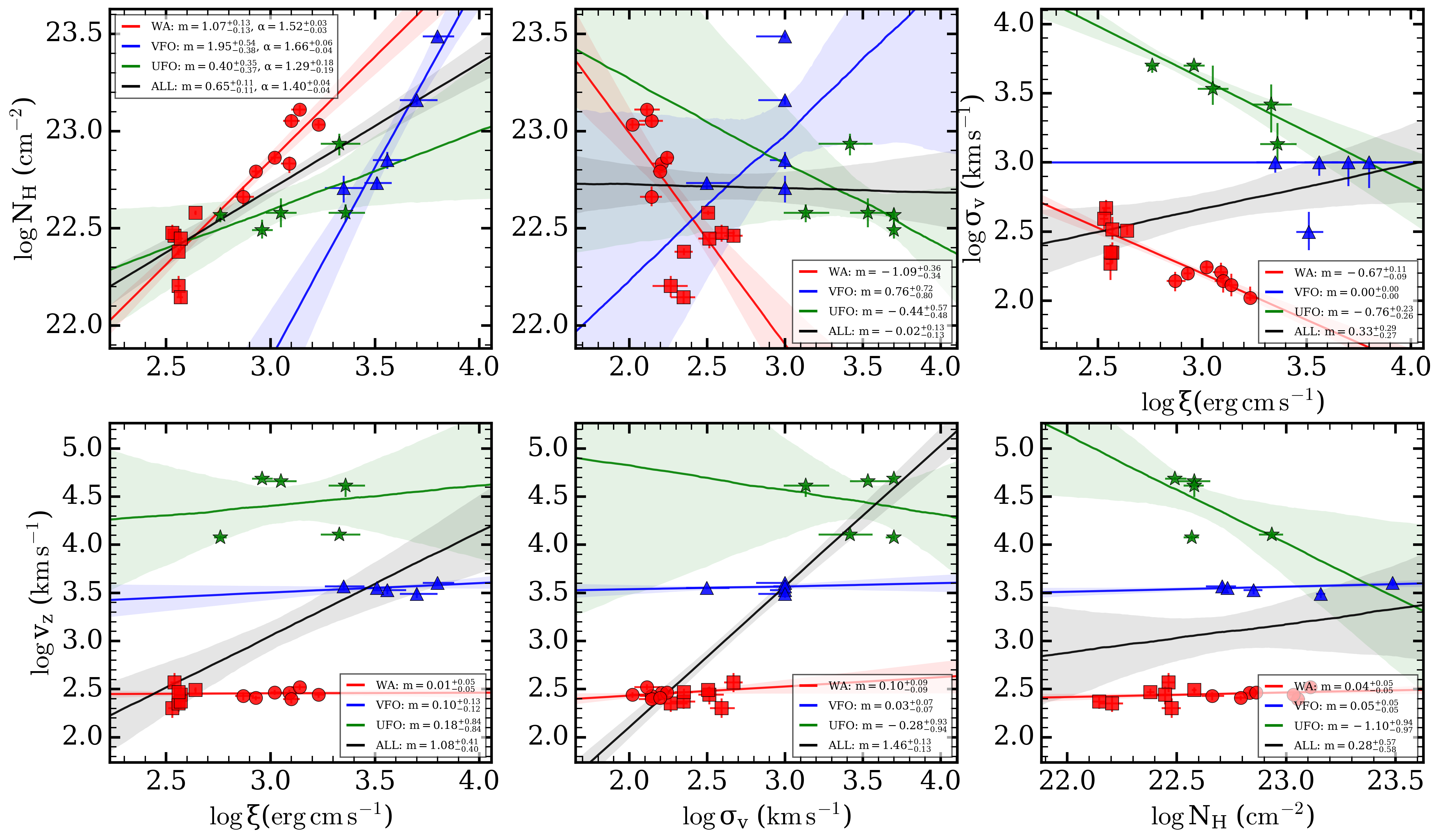}
    \caption{Pairwise relations among the fitted properties of all significant absorption components (\({\rm D.S.}>3\sigma\)) from the seven time-selected and HID-selected spectra analyzed in this work.  Circles, triangles, and stars denote WAs, VFOs, and UFOs, respectively. Colored lines show separate fits to each outflow class, while the black line shows the fit to the full combined sample; shaded bands mark the corresponding \(1\sigma\) uncertainties.}
    \label{fig:AMD}
\end{figure*}

To place the time- and HID-selected winds in the broader context of the full NGC~4151 outflow, Figure~\ref{fig:AMD} compares the properties of all significant absorption components (\({\rm D.S.}>3\sigma\)). The panels show pairwise relations among the fitted quantities \(N_{\rm H}\), \(\log\xi\), \(\sigma_v\), and \(v_z\), with the points grouped by outflow class (WA, VFO, and UFO). For each panel, we fit linear relations in log--log space using the \texttt{emcee} Markov Chain Monte Carlo sampler \citep{emcee_2013}, adopting a line model with intrinsic scatter and a split-normal approximation for asymmetric uncertainties. We use 48 walkers and 6000 steps per walker, and discard the first 1500 steps as burn-in. These fits are performed separately for each wind class, as well as for the full combined sample. 

\subsubsection{\(\log N_{\rm H}\) versus \(\log\xi\) }
The upper-left panel shows the absorption measure distribution (AMD) \citep{Holczer_2007, Behar_2009} for all wind components that are detected significantly (D.S. $> 3\sigma$), expressed here through the relation between \(\log N_{\rm H}\) and \(\log\xi\) . Following \citet{Behar_2009}, we parameterize the AMD as
\begin{equation}
    \log N_{\rm H} = m \log\xi + b ,
\end{equation}
where the slope \(m\) can be related to the wind density profile. For a large-scale outflow, the density profile is written as \(n(r)\propto r^{-\alpha}\), giving
\begin{equation}
    \alpha = \frac{1+2m}{1+m}.
\end{equation}
The fitted slopes are \(m=1.07\pm0.13\) for the WAs, \(m=1.95^{+0.54}_{-0.38}\) for the VFOs, \(m=0.40^{+0.35}_{-0.37}\) for the UFOs, and \(m=0.65\pm0.11\) for the full sample. These correspond to large-scale density profiles of \(\alpha=1.52\pm0.03\), \(1.66^{+0.06}_{-0.04}\), \(1.29^{+0.18}_{-0.19}\), and \(1.40\pm0.04\), respectively. Thus, the WA, VFO, and full-sample AMD slopes remain close to the \(n(r)\propto r^{-1.5}\) expectation of a Blandford--Payne-type magnetocentrifugal wind \citep{Blandford_Payne_1982}, while the VFOs are somewhat steeper. A similar steep AMD has also been reported for PG~1211+143, where \citealt{Reeves_2026} inferred \(a\simeq1.4\) and hence \(\alpha\simeq1.6\) for a large-scale wind. The UFO slope is flatter and more weakly constrained in the event-selected sample. Including the PV observations in Paper~I steepens the UFO AMD slope and brings it closer to the full-sample value.

\subsubsection{\(\log v_{\rm z}\) versus \(\log\xi\) }

In a smooth, large-scale wind, the AMD slope also predicts a relation between velocity and ionization. Since
\begin{equation}
    \xi = \frac{L}{n(r)r^2} \propto r^{\alpha-2},
\end{equation}
and for a magnetocentrifugal or Keplerian-like flow
\begin{equation}
    v_{\rm out} \propto r^{-1/2},
\end{equation}
one obtains
\begin{equation}
    v_{\rm out} \propto \xi^{1/[2(2-\alpha)]}.
\end{equation}
Thus, for \(\alpha=1.3-1.7\), the expected slope of \(\log v_{\rm out} -\log \xi\) is 0.7--1.6. By comparison, the lower-left panel shows that the within-class \(v_z\)--\(\xi\) trends are much flatter: the WAs are nearly flat, the VFOs show a moderate positive slope, and the UFOs remain shallow. This is the same basic discrepancy noted in Paper~I and also emphasized for PG~1211+143 \citep{Reeves_2026}, where the observed \(v\)--\(\xi\) slope implied a much flatter effective density profile than the AMD alone.

If instead the same AMD slope is interpreted in the small-scale limit \citep{Behar_2009}, where the ionization spread is driven primarily by local density variations or clump sizes at similar radius, the implied indices become much flatter, with \(\alpha_s= 1/(1+m) = 0.48\pm0.03\) for the WAs, \(0.34^{+0.06}_{-0.04}\) for the VFOs, \(0.71^{+0.18}_{-0.19}\) for the UFOs, and \(0.60\pm0.04\) for the full sample. In this case, the corresponding \(v_z\)--\(\xi\) dependence is of order \(v_z \propto \xi^{0.3}\), which is generally closer to the observed within-class trends than the steeper large-scale prediction. Meanwhile, the projection effects may also be important, as argued in Paper~I, where a bowl-shaped or curved streamline geometry can further flatten the observed \(v_z\)--\(\xi\) relation, especially for the more distant WA components. The data therefore favor a hybrid picture in which the wind is globally stratified, but shaped by both geometry and local clumpiness.

\subsubsection{\(\log N_{\rm H}\) versus \(\log \sigma_v\)}
The upper-middle panel of Figure~\ref{fig:AMD} shows \(\log N_{\rm H}\) versus \(\log \sigma_v\). The full-sample trend is nearly flat because the relation is dominated by the separation and limited sampling of the individual wind classes. The VFOs and UFOs are more weakly constrained in this panel. The WAs show some interesting internal structure. The square and circular symbols correspond to the two distinct WA components, and together they define a turnover-like pattern rather than a single monotonic sequence. The lower-ionization WA branch (squares) shows increasing \(N_{\rm H}\) with increasing \(\sigma_v\), while the higher-ionization WA branch (circles) shows the opposite tendency, with larger column densities occurring at smaller velocity widths. One possible interpretation is that the WA turnover reflects different layers of a slow, partially failed wind. In this picture, the two WA components need not trace a single monotonic \(N_{\rm H}\)--\(\sigma_v\) relation, because they may arise from different parts of the same circulating or stratified flow.

\subsubsection{\(\log v_z\) versus \(\log \sigma_v\)}
The lower-middle panel shows \(\log v_z\) versus \(\log\sigma_v\). The full sample shows a strong positive trend, driven primarily by the separation between WAs, VFOs, and UFOs. The within-class trends are much weaker. The positive class-to-class trend is qualitatively consistent with broader AGN-UFO demographic studies in which line width and outflow velocity are correlated \citep{Laurenti_2026}.

\subsubsection{\(\log \sigma_v\) versus \(\log \xi\)}
The upper-right panel shows \(\log\sigma_v\) versus \(\log\xi\). In the event-selected data, the WAs and UFOs both show negative trends, while the VFOs are essentially flat. When the PV observations are added, the VFO trend becomes more positive, but this behavior is not strongly required by the event-only sample. Thus, while broader components are generally associated with more extreme wind phases, \(\sigma_v\) is not determined by ionization alone and likely also depends on geometry, unresolved substructure, and local dynamical conditions. 

\subsubsection{\(\log v_z\) versus \(\log N_{\rm H}\)}
The lower-right panel shows \(\log v_z\) versus \(\log N_{\rm H}\). The combined sample shows a positive trend, indicating that faster components also tend to have larger columns. However, the within-class relations are shallow and even negative for UFOs, implying that this trend is driven primarily by the separation between the three wind classes rather than by a strong intrinsic scaling within each class.

\section{Summary and Discussion} \label{sec:discussion}
We analyzed seven integrated spectra constructed from fourteen XRISM/Resolve observations of NGC~4151, selected according to both local variability phases (Flare, Postflare, After, and Hard-dip) and global spectral hardness/intensity regimes (HS, LS, and LH). This analysis is organized around three related questions: \textit{when}, \textit{where}, and \textit{how} are the winds launched? 

The time- and HID-selected spectra address the first question, \textit{when}: all spectra require two persistent WA components, while the VFO and UFO phases are transient. The Postflare spectrum shows the richest structure, with two WAs, one VFO, and two UFOs, whereas the HS spectrum is adequately described by the WAs alone. This demonstrates that the fast wind is most visible after flare activity and in lower-flux, harder spectral regimes. We further find that the intrinsic ``hot'' obscurer is more strongly linked to the global hardness--intensity regime than to the short-timescale event selection. Its column density is anti-correlated with CINDICITY, indicating that brighter, softer spectra correspond to weaker intrinsic obscuration and weaker fast-wind absorption. The marginal positive relation between the intrinsic obscurer and the total wind column further suggests that the obscurer and ionized wind may respond to related changes in the inner source geometry.

The second question is \textit{where} the absorbing gas is located. The characteristic radius and volume filling factor constraints suggest that the WAs are unlikely to represent simple freely escaping outflows. They are instead more consistent with slow, stratified, partially failed, or circulating material. The derived energetics confirm that the UFOs are the dominant kinetic component of the outflow. In their nominal estimates, all of the UFO phases exceed the canonical $0.5\%\,L_{\rm Edd}$ feedback threshold. Under the most conservative assumptions, their kinetic powers remain close to this level, reaching marginal values of order $\sim 0.3\%\,L_{\rm Edd}$, so the UFOs remain promising candidates for feedback in NGC~4151 even if the threshold is not robustly exceeded in every case.

For the Postflare UFO and VFO components, the $\sim10$ ks response provides an additional compactness constraint. It corresponds to light-crossing distance of $r\sim c\Delta t\simeq60\,r_g$, which we interpret as the most compact causal limiting scale. If the Postflare VFOs and UFOs are associated with this compact scale, their fitted ionization parameters imply local densities of: $n\sim5.8\times10^{10}~{\rm cm^{-3}}$ and $f_v>1.8\times10^{-2}$ for the VFO, $n\sim1.8\times10^{11}~{\rm cm^{-3}}$ and $f_v>1.6\times10^{-3}$ for the slower UFO, and $n\sim3.4\times10^{11}~{\rm cm^{-3}}$ and $f_v>3.8\times10^{-4}$ for the faster UFO, requiring a clumpy or filamentary absorber. Under these compact-scale filling factors, the Postflare faster UFO reaches $\dot{E}_{\rm k}\lesssim0.2\% L_{\rm Edd}$. Alternatively, if the absorber thickness is comparable to the compact X-ray corona, $\Delta r \sim 10r_g$, combining the density, $n\sim N_H/\Delta r$ with the measured $L_{ion}/\xi$ places the Postflare VFO/UFO components at radii of $\sim 10^2-10^3$. The exact radius depends on the dominate cause of the $10$ ks flare--wind response.

The third question is \textit{how}: The AMD slope of the fastest outflows and their extreme outflow momentum rate are broadly consistent with a magnetically driven disk-wind structure. At the same time, the relatively slow, persistent WA components may be more strongly affected by radiative driving; their momentum requirements are less extreme than those of the fast transient components. The VFO/UFO phases appear to be closely connected to short-timescale coronal variability. Their preferential appearance in the Postflare, LS, and LH spectra suggests that the fastest winds may be triggered or made observable by flare-driven magnetic activity, possibly through magnetic reconnection or related coronal processes that lift dense gas from the inner disk.

In the following sections, we discuss the implications of these results for the disk-wind structure, geometry, variability, and launching physics, and then suggest future work.

\subsection{Disk Wind Structure and Geometry}
\subsubsection{Clumping}
The results suggest a globally organized but locally complex wind structure. The AMD trends imply a density profile of $n(r)\propto r^{-1.5}$ in the large-scale interpretation, broadly consistent with a magnetocentrifugal disk-wind structure \citep{Blandford_Payne_1982, Fukumura_2010, Fukumura_2015}. This is also consistent with the momentum argument discussed above: the VFOs and UFOs often carry momentum rates larger than the available radiative momentum flux, indicating that radiation pressure alone is unlikely to provide the main driving force. Thus, magnetic stresses may do much of the work in launching and accelerating the fastest phases.

MHD disk-wind models are intrinsically stratified in density, ionization, velocity, and opening angle, while local thermal/radiative instabilities can further break the flow into clumps or streams \citep{Fukumura_2010, Fukumura_2015,Dannen_Proga_2020,Waters_2021}. The flatter-than-expected $v_z-\xi$ relation indicates additional structure beyond a simple self-similar flow. In the small-scale interpretation of the AMD, the ionization spread is driven by local density gradients or clump sizes over short length scales, producing a much shallower effective scaling. This supports the idea that the fast outflows contain substantial small-scale substructure. Such clumpiness can arise through thermal instability in sub-Eddington outflows \citep{Dannen_Proga_2020, Waters_2021}. Radiation pressure may also contribute to shaping the gas locally. Even if it is not the dominant global driving mechanism, it can promote inhomogeneity, help structure the wind, and allow lower-ionization clumps to survive in regions where a smooth flow would otherwise be over-ionized \citep{Proga_Kallman_2004}.

A small-scale clumpy interpretation, however, is not the only explanation. As discussed in Paper~I, projection effects in a curved or bowl-like streamline geometry can further flatten the observed \(v_z\)--\(\xi\) relation, particularly for more distant components viewed at larger angles. Thus, the present results support an XRISM-era picture in which AGN winds are globally stratified and likely magnetically launched, but locally clumpy, geometrically complex, and strongly time dependent \citep{XRISMPDS456_2025, Reeves_2026, Gu_2025, Xiang_2025}.

\subsubsection{Failed WA and Re-emission}
The WAs are difficult to interpret as freely escaping gas based on their outflow radius constraints. Instead, they may trace the gas that is lifted from the disk, later stalls, and circulates, and may typically be observed far from its launching point. This interpretation is consistent with \citealt{Miller_2026}. The persistent X-ray WAs are also broadly consistent with the long-known UV/X-ray absorbing outflow structure in NGC~4151, in the sense that both trace slow, multi-zone gas with velocities of a few hundred to \(\sim10^3~{\rm km~s^{-1}}\) and line widths of FWHM $\sim 90-900~\mathrm{km~s^{-1}}$ \citep{Crenshaw_Kraemer_2012, Couto_2016}. We note, however, that the X-ray WAs in NGC~4151 are too hot and too highly ionized to be identified with the dusty failed winds often invoked in BLR-formation models \citep{Czerny_2017, Matthews_2020}. The turnover-like behavior of the two WA branches in the $N_H -\sigma_v$ plane may reflect different layers of this slow, stratified, partially failed flow. 

A complementary constraint on the volume filling factor can be estimated from the WA re-emission ($pion^{emis\#2}$). Use the emission broadening ($\sigma_{\rm vgau}$), we can determine the Keplerian radius, $r_k = (c/\sigma_{\rm vgau})^2 r_g$, the density ($n=L_{\rm ion}/\xi r_k^2$), and the filling factor $f_v = \Delta r/r_k = N_H/nr_k$. This gives physically plausible filling factors only for the initial post-flare and LS spectra, with $f_v \sim 1.1$ and $f_v \sim 0.46$ respectively. Including the fitted emission covering factors gives an effective global filling factor of $\Omega f_v \sim 0.2$ in both cases. However, \citealt{Miller_2026} found a much larger emission covering factor close to unity and a radial filling factor of $\Delta r/r \sim 0.6$. This contrasts with the 0.9 Ms stacked spectrum, which is informative. The time-averaged WA emission could sample a broad, extended region of the WAs, while the time-resolved spectra with smaller effective covering factors could suggest that this re-emitting gas is not equally visible in every variability phase and may hint at geometric variability in the WAs.

In the remaining spectral windows, the same calculation gives either unphysical $f_v > 1$ values or is poorly constrained because the emission broadening is consistent with zero. This is likely due to the strong sensitivity required for the emission width since $f_v \propto r_k \propto \sigma_{\rm vgau}^{-2}$. The cases with unphysical filling factor indicate that the one-zone Keplerian-broadening estimation is unreliable when the WA re-emission is weak, phase-dependent, or geometrically complex.

\subsubsection{The Transient Re-emission of UFO}
Emission from the slower UFO ($pion^{\rm emis\#4}$) is only significant in the Postflare window, indicating a flare-related evolution. The transient visibility may reflect that its detection requires particular gas properties and positioning relative to the line-of-sight. 

Similar transient or flare-related Fe-K emission has been reported in other AGN, for example NGC 3516 \citep{Iwasawa_2004}, Mrk 766 \citep{Pounds_2003}, and PG1211+143 \citep{Pounds_Reeves_2009}, where variable Fe-K emission features were interpreted in terms of localized flare illumination, transient ejecta, or wide-angle wind re-emission. 

The re-emission in the Postflare UFO has a modest covering factor of $\Omega \sim 0.1$, suggesting that the emitting region of the slower UFO has a narrow solid angle, or is only visible over a small range of the streamlines during the post-flare window. In addition, it does not require additional Gaussian broadening ($\sigma_{\rm vgau} = 0^{+642}_{-0}~\mathrm{km~s^{-1}}$).  Using the upper limit on broadening gives a lower limit on the Keplerian radius of $r_K > 2\times10^5 r_g$, while the light crossing-timescale indicates a radius of order $\sim 60 r_g$. The discrepancy suggests that the absorption and re-emission may not trace the same part of the flow. The absorption is produced by a compact, denser, faster LOS clump closer to the BH, while the emission is produced by a wider-angle, more extended, and lower-velocity part of the same gas at a larger radius.

The variability timescale also supports the visibility interpretation rather than the physical disappearance of the gas. At $\sim 60~r_g$, the flow time for gas moving at $\sim 10^4~\rm{km/s}$ is $\sim3.5$ days, and the local orbital time ($\sim 2\pi (r/r_g)^{3/2}GM/c^3$) is $\sim5$ days. Therefore, the absence of this re-emission in the Flares and Afters spectra cannot mean that the entire emitting structure forms and disappears within $\sim 10$ ks. Instead, the gas likely persists, but its observability changes rapidly due to the combination of changes in ionization state, column density, illumination, and location. The Postflare spectrum may therefore capture a short-lived window in which the slower UFO is both strongly intersecting our LOS and sufficiently visible over a wider solid angle to reveal its emission counterpart.

\subsubsection{The Blue Shifted Emissions}
In addition to the linked UFO re-emission in the Postflare spectrum, several spectra require a separate blue-shifted emission component, $pion^{\rm{blue-emis}}$. This component is not linked to a specific LOS absorber and appears to have a different origin. It is not required in the flare or Postflare spectra, but is detected in the After spectrum and in all three integrated HID-selected spectra. This behavior suggests that $pion^{\rm blue-emis}$ is not simply the immediate re-emission counterpart of the Postflare UFO absorber. Instead, it may trace gas that has moved out of the direct line of sight after the flare/postflare phase, or a more persistent off-axis Fe-K emitting structure that becomes easier to detect in the more normal or time-averaged spectra. One possible interpretation is that the flare perturbs the inner disk atmosphere, compressing or lifting dense clumps while radiation pressure helps shape their ionization and column density. Magnetocentrifugal forces may then accelerate some of this material upward from the disk. When the clump or stream crosses our line of sight during the Postflare phase, it appears primarily as VFO/UFO absorption; after it moves out of the line of sight, the same or related material may still be visible through blue-shifted Fe-K emission. This interpretation is broadly consistent with recent XRISM results on NGC~1068, where broad Fe~XXV and Fe~XXVI emission is interpreted as a highly ionized, kinematically distinct phase of a bipolar outflow rather than as emission directly tied to a single line-of-sight absorber \citep{Bianchi_2026}.

If interpreted as wind emission,  $pion^{\rm{blue-emis}}$ would have velocities in $\sim10^4~\rm{km~s^{-1}}$ range and ionization parameters of $\log\xi \sim 2.6-2.7$, primarily producing Fe~XXIII -- Fe~XXV, with minor contribution from Fe~XXVI. The ionization state of the emission is lower than that of the fastest UFO absorbers; it may trace an off-axis part of the wind moving at VFO or slow-UFO velocities.  In Paper~1, this component was already interpreted as evidence for asymmetric wind emission associated with the BLR and failed gas streams on the far side of the disk. The net blueshift would imply that the approaching side above our LOC or the failed side at the far side of the disk dominates the observed emission, while the receding side is weaker, obscured, or occulted by the disk and inner absorber.

However, the wind-emission interpretation is not unique. Because the feature lies in the Fe-K band and is most evident in the after and integrated state spectra, it may instead represent part of the Fe-K emission-line structure from the inner accretion disk, such as the blue wing of a relativistically broadened reflection component. The 0.9 Ms XRISM/Resolve spectrum of NGC~4151 requires broad Fe-K emission that can be modeled with inner-disk reflection \citep{Miller_2026}. In the present event-resolved spectra, we model the feature phenomenologically as blue-shifted wind emission because of its centroid, ionization structure, and phase dependence. However, a full test of the degeneracy between wind emission and relativistic reflection is beyond the scope of this paper and is deferred to future work.

\subsection{The full picture of the wind variation and threshold feedback}
\begin{figure*}
    \centering
    \includegraphics[width=\linewidth]{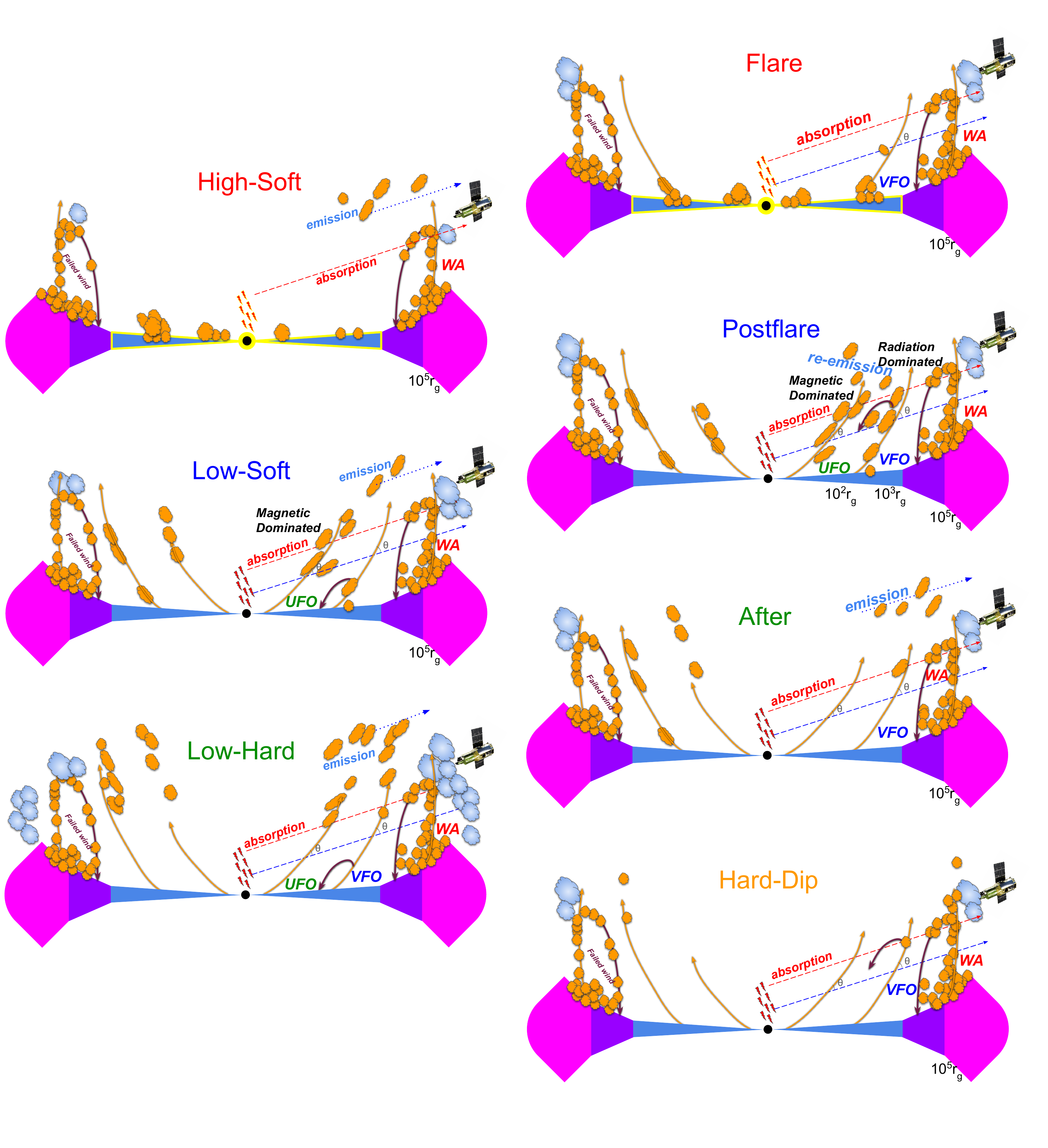}
    \caption{Schematic illustration of the proposed wind geometry and visibility windows in NGC~4151. The WAs form an extended, partially failed or circulating component at large radii and are present in all selections, while the VFO and UFO phases are more transient and depend on both spectral state and line-of-sight geometry. In the HS state, the source is bright and soft, and the fast wind is weak or absent. In the Postflare phase, roughly \(10\) ks after the flare selection, the VFO/UFO components become most visible, and the slower UFO can produce transient re-emission. In the LS and LH states, the lower continuum level favors stronger fast-wind absorption, with the LH state providing the clearest window for VFOs. The schematic emphasizes that the time-selected spectra probe short-timescale line-of-sight visibility, while the HID-selected spectra probe the global accretion-state dependence of the wind.
    }
    \label{fig:catoon}
\end{figure*}

%Summarize the picture with figure \ref{fig:catoon} 
The full pictures of the wind variation revealed by the time- and HID- selected spectra could be summarized in Figure \ref{fig:catoon}. In the HS state, the line of sight primarily intersects the WAs, while the fast VFO/UFO phases are absent. The blue gas in the schematic represents the intrinsic ``hot'' obscurer, whose changes are most clearly associated with the global state selections. This obscurer is weakest in the bright, soft HS state and becomes strongest in the Low-Hard state. During the flares, the ionizing continuum reaches its highest level, but the fast wind is not yet at its strongest. The gas may still be responding to the continuum change, over-ionized, or geometrically not blocking our LOS to the X-ray source. Roughly 10 ks later, in the Postflare phase, the VFO and UFO absorption strengthen, and the slower UFO briefly appears in both absorption and re-emission. By the After phase, this transient UFO re-emission has faded, while the blue-shifted Fe-K emission from the broader wind structure remains visible. In the hard-dip spectra, the remaining visible VFO component does not develop into an escaping outflow. The LS and LH spectra both favor UFO absorption, suggesting that lower-flux states provide a broader global window for detecting the fast wind. 

%Discuss the launching window of fast outflows, and compare it with NGC 3783 cite Liyi.
We want to emphasize that the local event can occur within any global state. The Flare, Postflare, After, and Hard-dip spectra isolate short-timescale changes in LOS visibility and ionization response, while the HS, LS, and LH spectra describe the broader accretion-state dependence of the outflow. In this schematic framework, the Postflare enhancement indicates a delayed local response to continuum variability. A similar flare-related picture has recently been reported in NGC 3783 \citep{Gu_2025}, where a launch of a UFO is synchronized with the sharp decay within $\sim 50$ ks. By analogy, if the $\sim 10$ ks delay is interpreted as a travel or crossing time for the Postflare UFOs, the slower UFO $pion_{\#4}$ would move only $\sim 2.6~r_g$, while the faster UFO $pion_{\#5}$ would move $\sim 10~r_g$. These distances are comparable to plausible coronal scales \citep{Fabian_2015}. Hence, the Postflare delay may trace the time for the newly accelerated UFOs to cross the compact X-ray source along our line of sight. 

%Discuss the frequency of ``threshold feedback" winds in sub-eddington source. Compare it with high accretion source such as PDS 456 and PG1211.
In the nominal values of the kinetic energetics, the UFOs exceed the $0.5\%~L_{\rm{Edd}}$ threshold feedback, and even under the most conservative filling-factor assumptions, they remain close to this level. This suggests that the feedback-level winds in sub-Eddington NGC 4151 are intermittent rather than continuous. This differs from near or super-Eddington systems such as PDS 456 and PG 1211+143, where powerful, structured UFOs appear to be a more persistent feature \citep{XRISMPDS456_2025, Reeves_2026}. The feedback picture in NGC 4151 implies that sub-Eddington Seyferts can occasionally access the feedback-level wind, but with a lower duty cycle and stronger dependence on spectral state and LOS geometry.

\subsection{Suggested Future Work}
The time- and HID-selected spectra presented in this work provide a first view of \textit{when} different wind phases in NGC 4151 become visible, but several caveats remain. 

The spectra are modeled separately. Future work should model these spectra with fully time-dependent photoionization and dynamical wind calculations (e.g., TPHO \citep{Rogantini_2022}, TEPID \citep{Luminari_2023}). The transient UFO detections in the Postflare, LS, and LH spectra may not only be windows of observability but may reflect real physical evolution of the gas. For instance, after flare-driven launching or lifting from the disk, the outflow may expand, lower its density, and become more highly ionized, causing the Fe-K absorption to weaken or disappear. 

We also assume a static FUV/UV contribution to the ionizing SED. Since the UV-emitting disk is expected to vary more slowly than the X-ray continuum, this approximation is unlikely to dominate the short-timescale wind variability discussed here, but future analysis on simultaneous UV/X-ray monitoring would be useful for quantifying any SED-driven changes in the inferred ionization parameters. Such observations would also test whether any UV absorption counterpart responds during the same postflare or low-flux windows in which the fast X-ray outflows are strongest. 

In addition, several key components are potentially degenerate. The variable ``hot'' obscuration can mimic continuum-state changes due to lack of data in the soft band, wind emission can overlap with relativistic Fe-K reflection, and the derived energetics depend on uncertain volume filling factors and covering geometry.

A natural next step is a dedicated variability and lag analysis between the wind components, the continuum source, and the intrinsic obscurer. The Postflare enhancement of the fast wind could reflect recombination, changing illumination, or the motion of clumpy gas across the line of sight, but these possibilities cannot be separated from the spectra alone. Time-lag measurements would test whether the fast outflows respond causally to flares and whether transient absorption modifies the observed continuum reverberation signal. This is especially important for complex or CLAGN, where apparent lags may be shaped by both intrinsic accretion variability and evolving LOS absorption \citep{Feng_2024}.

A useful extension of this work would be to develop a CINDICITY-like convention that can be applied consistently across different AGN and instruments. In this paper, CINDICITY is defined using the count rate and hardness ratio within the XRISM/Resolve band and is therefore optimized for tracking the spectral evolution of NGC~4151 in this specific dataset. A flux-defined hardness ratio would help direct comparison with other sources or with instruments.

%relativistic reflection degeneracy with winds
In addition, resolving the degeneracy between wind emission, absorption, and relativistic reflection is important. In this work, the blue-shifted Fe-K emission component is modeled phenomenologically as wind-related emission, but part of the same Fe-K excess could instead arise from the blue wing of relativistic disk reflection \citep{Miller_2026}. Future work should fit the Fe-K excess in the event/state-resolved spectra with a self-consistent reflection model to separate disk reflection from wind-related Fe-K emission.

%Jets? Radio observation? connecting jet with wind
Future observations should also test whether the wind variability is connected to jet or radio activity. The tendency for the VFOs to appear most strongly in harder, lower-flux selections raises the possibility that the wind and jet are coupled through changes in the inner accretion flow. The results of AMD and momentum arguments point toward a magnetic contribution to the fast wind. Relevantly, VLA polarization observations of NGC 4151 reveal stratified magnetic-field structures in the radio outflow \citep{Ghosh_2026}. More broadly, an inverse relation between wind column and radio loudness is observed in radio-loud AGN, which was interpreted as a change in magnetic-field configuration powering either wind or jet \citep{Mehdipour_2019}. Coordinated XRISM and high-resolution radio monitoring of NGC 4151 could therefore test whether the strongest fast outflow phases coincide with changes in radio emission.

\section{Conclusion} \label{sec:summary}
We performed a time-resolved study of disk wind variability using fourteen XRISM observations of NGC 4151. By combining spectral hardness/intensity classification (HS, LS, LH) with local variability phases (Flares, Postflares, Afters, and Hard-dips), we explored the wind variability in response to the central engine. Our main conclusions are:

\begin{itemize}
    \item The wind in NGC~4151 is strongly time- and hardness/intensity-dependent. All spectra require two persistent WA components, while the faster VFO and UFO phases are transient. The Postflare spectrum shows the richest wind structure, with two WAs, one VFO, and two UFOs, whereas no fast wind is detected in the HS state and the spectrum is dominated by WAs only. This indicates that the fast wind is most visible after flare events and in lower-flux states. The Postflare UFO and VFO may be a physical consequence of magnetic/coronal activity, becoming observable as the flare decays, while the lower-flux states may instead be partly shaped by enhanced absorption or obscuration.

    \item The intrinsic ``hot'' obscurer is closely tied to the global spectral state. Its column density is lowest in the bright-soft HS state and highest in the LH state, and it is anti-correlated with CINDICITY. The marginal correlation between the intrinsic obscurer column and the total wind column further suggests that the obscurer and the ionized wind may respond to related changes in the inner source geometry.

    \item The strongest transient wind signatures tend to appear during the Postflare phases rather than at the flare peak, suggesting that the fast outflows are a physical consequence of the flaring activity. In this picture, the UFOs become most observable as the flare decays, possibly after magnetic activity in the corona has launched or lifted dense gas from the inner disk. For the black-hole mass of NGC~4151, a timescale of $\sim10$ ks corresponds to the light-crossing time of $\sim60\,r_g$, indicating that the relevant variability is associated with compact inner-disk/coronal scales. If the same timescale is interpreted as a vertical or source-crossing time, the observed velocities imply that Postflare UFOs move by distances comparable to plausible compact coronal scales. Thus, the Postflare spectrum likely captures a short-lived phase in which flare drives inner-disk activity and produces observable fast outflows.

    \item The radius and filling-factor constraints indicate that the WAs are unlikely to be simple freely escaping winds. Instead, they are more consistent with a slow, stratified, partially failed or circulating component. The WA-linked re-emission gives physically plausible filling factors only in selected spectra, suggesting that the re-emitting WA gas is globally present but not equally visible in every variability phase.

    \item The UFOs dominate the kinetic power of the outflow. In the nominal estimates, all UFO phases exceed the $0.5\%\,L_{\rm Edd}$ feedback threshold, while under the most conservative filling-factor corrections they remain marginally close to this level, at values of order $\sim0.3\%\,L_{\rm Edd}$. This suggests that feedback-level winds in sub-Eddington NGC~4151 are intermittent rather than continuous.

    \item The AMD and momentum-rate arguments favor a hybrid launching picture. The AMD is broadly consistent with a large-scale, stratified, magnetocentrifugal wind, while the large VFO/UFO momentum rates imply that radiation pressure alone is unlikely to explain the fastest phases. At the same time, the flattened $v_z$--$\xi$ relations, transient re-emission, and blue-shifted Fe-K emission point to a wind that is locally clumpy, geometrically complex, and strongly phase dependent.

\end{itemize}

\begin{deluxetable*}{cclllllll} 
\tabletypesize{\footnotesize}
\label{table:parameters}
\tablecaption{Best-fit Parameters and Detection Significance (D.S) for the outflow components.}
\tablewidth{0pt}
\tablehead{
\colhead{Components} & \colhead{Parameters} & \colhead{Flares} & \colhead{Postflares} & \colhead{Afters} & \colhead{Hard-Dips} & \colhead{HS} & \colhead{LS} & \colhead{LH}
}
\startdata
    $pion_{\#1}$ & $N_H$ ($10^{22}~\rm cm^{-2}$) 
        & $2.9^{+0.2}_{-0.2}$ & $3.0^{+0.3}_{-0.3}$ & $2.8^{+0.2}_{-0.3}$ & $1.6^{+0.2}_{-0.2}$ & $1.4^{+0.1}_{-0.1}$ & $2.4^{+0.1}_{-0.1}$ & $3.8^{+0.1}_{-0.1}$\\
        & $\log \xi$ ($\mathrm{erg~cm~s^{-1}}$) 
        & $2.54^{+0.02}_{-0.02}$ & $2.53^{+0.02}_{-0.02}$ & $2.57^{+0.01}_{-0.01}$ & $2.56^{+0.01}_{-0.02}$ & $2.57^{+0.01}_{-0.02}$ & $2.56^{+0.01}_{-0.01}$ & $2.64^{+0.01}_{-0.01}$\\
        & $\sigma_v$ ($\mathrm{km~s^{-1}}$) 
        & $470^{+70}_{-60}$ & $390^{+80}_{-60}$ & $330^{+80}_{-50}$ & $190^{+50}_{-40}$ & $220^{+40}_{-40}$ & $220^{+30}_{-30}$ & $320^{+30}_{-20}$\\
        & $v_z$ ($\mathrm{km~s^{-1}}$) 
        & $-370^{+90}_{-100}$ & $-200^{+90}_{-80}$ & $-280^{+60}_{-70}$ & $-230^{+40}_{-40}$ & $-230^{+40}_{-40}$ & $-290^{+30}_{-30}$ & $-310^{+20}_{-20}$\\
    $\Delta C/\Delta \nu$ & & -102/-4 & -74/-4 & -93/-4 & -63/-4 & -106/-4 & -242/-4 & -155/-4\\
    D.S ($\Delta$AIC) & & $\infty$ (-93) & 7.9$\sigma$ (-64) & $\infty$ (-84) & 7.2$\sigma$ (-53) & $\infty$ (-97) & $\infty$ (-287) & $\infty$ (-117)\\
    \hline
    $pion_{\#2}$ & $N_H$ ($10^{22}~\rm cm^{-2}$) 
        & $6.8^{+0.6}_{-0.7}$ & $4.6^{+0.6}_{-0.5}$ & $11^{+1}_{-1}$ & $13^{+1}_{-1}$ & $7.3^{+0.5}_{-0.4}$ & $6.2^{+0.4}_{-0.4}$ & $11^{+1}_{-1}$\\
        & $\log \xi$ ($\mathrm{erg~cm~s^{-1}}$) 
        & $3.09^{+0.03}_{-0.04}$ & $2.87^{+0.05}_{-0.04}$ & $3.10^{+0.04}_{-0.04}$ & $3.14^{+0.03}_{-0.04}$ & $3.02^{+0.03}_{-0.02}$ & $2.93^{+0.02}_{-0.02}$ & $3.23^{+0.03}_{-0.02}$\\
        & $\sigma_v$ ($\mathrm{km~s^{-1}}$) 
        & $160^{+30}_{-20}$ & $140^{+20}_{-20}$ & $140^{+30}_{-30}$ & $130^{+30}_{-30}$ & $180^{+20}_{-20}$ & $160^{+20}_{-20}$ & $110^{+30}_{-5}$\\
        & $v_z$ ($\mathrm{km~s^{-1}}$) 
        & $-290^{+20}_{-20}$ & $-270^{+30}_{-30}$ & $-250^{+20}_{-30}$ & $-330^{+20}_{-20}$ & $-290^{+20}_{-20}$ & $-260^{+20}_{-20}$ & $-280^{+20}_{-20}$\\
    $pion^{emis\#2}$ & $v_z$ ($\mathrm{km~s^{-1}}$) 
        & $570^{+130}_{-100}$ & $1000^{+0}_{-110}$ & $580^{+70}_{-70}$ & $460^{+50}_{-60}$ & $560^{+60}_{-70}$ & $880^{+120}_{-110}$ & $620^{+90}_{-130}$\\
        & $\Omega_{\rm emis}$ ($4\pi$) 
        & $0.07^{+0.02}_{-0.02}$ & $0.19^{+0.04}_{-0.04}$ & $0.18^{+0.02}_{-0.03}$ & $0.08^{+0.02}_{-0.02}$ & $0.11^{+0.02}_{-0.02}$ & $0.38^{+0.03}_{-0.02}$ & $0.10^{+0.03}_{-0.02}$\\
    $vgau$ & $\sigma$ ($\mathrm{km~s^{-1}}$) 
        & $0^{+1500}_{-0}$ & $640^{+270}_{-190}$ & $270^{+60}_{-60}$ & $0^{+230}_{-0}$ & $200^{+80}_{-80}$ & $1220^{+110}_{-100}$ & $280^{+140}_{-80}$\\
    $\Delta C/\Delta \nu$ & & -135/-7 & -100/-7 & -140/-7 & -89/-7 & -225/-7 & -303/-7 & -133/-7\\
    D.S ($\Delta$AIC) & & $\infty$ (-119) & $\infty$ (-83)  & $\infty$ (-125) & 8.2$\sigma$ (-73) & $\infty$ (-210) & $\infty$ (-287) & $\infty$ (-117)\\
    \hline
    $pion_{\#3}$ & $N_H$ ($10^{22}~\rm cm^{-2}$) 
        & $7.1^{+0.7}_{-0.7}$ & $31^{+1}_{-1}$ & $5.1^{+0.8}_{-0.8}$ & $14^{+1}_{-1c }$ & $3.3^{+1.3}_{-1.2}$ & $0.1^{+0.1}_{-0.1}$ & $5.4^{+0.4}_{-0.4}$\\
        & $\log \xi$ ($\mathrm{erg~cm~s^{-1}}$) 
        & $3.56^{+0.09}_{-0.07}$ & $3.80^{+0.08}_{-0.07}$ & $3.35^{+0.10}_{-0.09}$ & $3.70^{+0.10}_{-0.08}$ & $3.60^-$ & $3.60^-$ & $3.51^{+0.07}_{-0.06}$\\
        & $\sigma_v$ ($\mathrm{km~s^{-1}}$) 
        & $1000^{+0}_{-200}$ & $1000^{+0}_{-580}$ & $1000^{+0}_{-160}$ & $1000^{+0}_{-330}$ & $900^-$ & $900^-$ & $320^{+120}_{-80}$\\
        & $v_z$ ($\mathrm{km~s^{-1}}$) 
        & $-3400^{+430}_{-440}$ & $-4000^{+300}_{-300}$ & $-3700^{+500}_{-500}$ & $-3100^{+450}_{-440}$ & $3500^-$ & $3500^-$ & $-3600^{+120}_{-110}$\\
    $\Delta C/\Delta \nu$ & & -20/-4 & -18/-4 & -20/-4 & -17/-4 & -6/4$^*$ &  -0/4$^*$ & -44/-4\\
    D.S ($\Delta$AIC) & & 3.4$\sigma$ (-10) & 3.3$\sigma$ (-9) & 3.5$\sigma$ (-20) & 3.1$\sigma$ (-8) & $\leq 1\sigma$ (3) & $\leq 1\sigma$ (10) & 5.8$\sigma$ (-35)\\
    \hline
     $pion_{\#4}$ & $N_H$ ($10^{22}~\rm cm^{-2}$) 
        & $4.4^{+0.7}_{-0.6}$ & $8.6^{+1.1}_{-1.1}$ & $2.2^{+1.0}_{-1.0}$ & $2.2^{+0.6}_{-0.6}$ & $0.5^{+0.5}_{-0.5}$ & $3.7^{+0.3}_{-0.3}$ & $0.7^{+0.3}_{-0.2}$\\
        & $\log \xi$ ($\mathrm{erg~cm~s^{-1}}$) 
        & $3.64^{+0.17}_{-0.68}$ & $3.33^{+0.10}_{-0.09}$ & $3.20^-$ & $3.38^{+0.13}_{-0.11}$ & $3.20^-$ & $2.76^{+0.03}_{-0.03}$ & $2.91^{+0.10}_{-0.10}$\\
        & $\sigma_v$ ($\mathrm{km~s^{-1}}$) 
        & $260^{+4700}_{-160}$ & $2600^{+1000}_{-970}$ & $1800^-$ & $100^{+70}_{-0}$ & $1800^-$ & $5000^{+0}_{-550}$ & $1300^{+430}_{-320}$\\
        & $v_z$ ($\mathrm{km~s^{-1}}$) 
        & $-10000^{+290}_{-1900}$ & $-13000^{+830}_{-980}$ & $-15000^-$ & $-13600^{+90}_{-80}$ & $-15000^-$ & $-12000^{+480}_{-620}$ & $-27200^{+580}_{-500}$\\
    $pion^{emis\#4}$ & $v_z$ ($\mathrm{km~s^{-1}}$) 
        & ... & $-9000^{+200}_{-100}$ & ... & ... & ... & ... & ...\\
        & $\Omega_{\rm emis}$ ($4\pi$) 
        & ... & $0.10^{+0.06}_{-0.04}$ & ... & ... & ... & ... & ...\\
    $vgau$ & $\sigma$ ($\mathrm{km~s^{-1}}$) 
        & ... & $0^{+640}_{-0}$ & ... & ... & ... & ... & ...\\
    $\Delta C/\Delta \nu$ & & -7/-4 & -26/-7 & -5/-4$^*$ & -9/-4 & 0.3/-4$^*$ & -114/-4 & -14/-4\\
    D.S ($\Delta$AIC) & & 1.6$\sigma$ (1.5) & 3.4$\sigma$ (-9) & $\leq 1\sigma$ (4) & 1.8$\sigma$ (0.4) & $\leq 1\sigma$ (8) & $\infty$ (-105) &  2.7$\sigma$ (-4.5)\\
    \hline
    $pion_{\#5}$ & $N_H$ ($10^{22}~\rm cm^{-2}$) 
        & $1.5^{+0.7}_{-0.7}$ & $3.8^{+0.7}_{-0.6}$ & $2.3^{+1.0}_{-1.0}$ & $1.1^{+0.8}_{-0.9}$ & $1.1^{+0.5}_{-0.5}$ & $3.1^{+0.4}_{-0.3}$ & $3.8^{+0.4}_{-0.4}$\\
        & $\log \xi$ ($\mathrm{erg~cm~s^{-1}}$) 
        & $3.10^-$ & $3.05^{+0.08}_{-0.07}$ & $3.10^-$ & $3.10^-$ & $3.10^-$ & $2.96^{+0.05}_{-0.05}$ & $3.36^{+0.09}_{-0.08}$\\
        & $\sigma_v$ ($\mathrm{km~s^{-1}}$) 
        & $3300^-$ & $3400^{+1600}_{-800}$ & $3300^-$ & $3300^-$ & $3300^-$ & $5000^{+0}_{-350}$ & $1400^{+570}_{-380}$\\
        & $v_z$ ($\mathrm{km~s^{-1}}$) 
        & $-45000^-$ & $-45900^{+1400}_{-1100}$ & $-45000^-$ & $-45000^-$ & ... & $-48700^{+1900}_{-1500}$ & $-41300^{+440}_{-9800}$\\
    $\Delta C/\Delta \nu$ & & -4/-4$^{*}$ & -20/-4 & -5/-4$^{*}$ & -2/-4$^{*}$ & -4/-4$^{*}$ & -33/-4 & -21/-4\\
    D.S ($\Delta$AIC) & & $\leq 1\sigma$ (4)  & 3.6$\sigma$ (-11) & $\leq 1\sigma$ (3) & $\leq 1\sigma$ (7) & $\leq 1\sigma$ (4) & 4.8$\sigma$ (-24) & 3.6$\sigma$ (-12)\\
    \hline
    $pion^{blue-emis}$ & $\log \xi$ ($\mathrm{erg~cm~s^{-1}}$) 
        & ... & ... & $2.74^{+0.07}_{-0.05}$ & ... & $2.64^{+0.08}_{-0.02}$ & $2.64^{+0.02}_{-0.02}$ & $2.58^{+0.02}_{-0.02}$\\
        & $v_z$ ($\mathrm{km~s^{-1}}$) 
        & ... & ... & $-7400^{+60}_{-60}$ & ... & $-10000^{+1900}_{-0}$ & $-8200^{+240}_{-330}$ & $-11700^{+220}_{-220}$\\
        & $\Omega$ 
        & ... & ... & $0.044^{+0.011}_{-0.010}$ & ... & $0.080^{+0.014}_{-0.010}$ & $0.074^{+0.009}_{-0.008}$ & $0.071^{+0.009}_{-0.009}$\\
    $vgau$ & $\sigma$ ($\mathrm{km~s^{-1}}$) 
        & ... & ... & $0^{+130}_{-0}$ & ... & $2040^{+1050}_{-630}$ & $710^{+360}_{-180}$ & $540^{+210}_{-120}$\\
    $\Delta C/\Delta \nu$ & & ... & ... & -18/-4 & ... & -54/-4 & -27/-4 & -59/-4\\
    D.S ($\Delta$AIC) & & ... & ... & 3.2$\sigma$ (-9) & ... & 6.6$\sigma$ (-45) & 4.3$\sigma$ (-18) & 6.9$\sigma$ (-50)\\
    \hline
    $pow$ & $\Gamma$ 
        & $1.77^{+0.02}_{-0.01}$ & $1.68^{+0.02}_{-0.01}$ & $1.69^{+0.03}_{-0.01}$ & $1.71^{+0.01}_{-0.02}$ & $1.70^{+0.01}_{-0.02}$ & $1.73^{+0.03}_{-0.05}$ & $1.68^{+0.02}_{-0.01}$\\
        & $\Delta \Gamma$ 
        & $-0.35^{+0.02}_{-0.07}$ & $-0.26^{+0.02}_{-0.10}$ & $-0.33^{+0.02}_{-0.09}$ & $-0.37^{+0.06}_{-0.06}$ & $-0.32^{+0.03}_{-0.04}$ & $-0.47^{+0.12}_{-0.10}$ & $-0.31^{+0.02}_{-0.10}$\\ 
        & $E_0$ 
        & $9.16^{+0.36}_{-0.32}$ & $9.7^{+1.1}_{-0.6}$ & $9.6^{+0.4}_{-0.7}$ & $9.3^{+0.6}_{-0.4}$ & $9.1^{+0.2}_{-0.5}$ & $9.8^{+1.5}_{-0.7}$ & $8.6^{+0.3}_{-0.6}$\\ 
        & $\mathrm{Norm}~\times 10^7$ 
        & $2.89^{+0.09}_{-0.09}$ & $2.47^{+0.11}_{-0.10}$ & $2.2^{+0.2}_{-0.9}$ & $2.35^{+0.07}_{-0.10}$ & $2.39^{+0.04}_{-0.09}$ & $2.37^{+0.17}_{-0.26}$ & $2.03^{+0.08}_{-0.08}$\\ 
    \hline
    $hot$ & $N_H$ ($10^{22}~\rm cm^{-2}$) 
        & $17.3^{+0.4}_{-0.7}$ & $17.5^{+0.8}_{-0.9}$ & $17.9^{+0.8}_{-1.2}$ & $18.0^{+0.8}_{-0.7}$ & $12.9^{+0.2}_{-0.9}$ & $17.8^{+2.0}_{-1.5}$ & $20.1^{+0.5}_{-1.3}$\\
        & $f_{\rm cov}$ 
        & $0.857^{+0.007}_{-0.007}$ & $0.847^{+0.013}_{-0.006}$ & $0.859^{+0.014}_{-0.008}$ & $0.857^{+0.007}_{-0.008}$ & $0.846^{+0.015}_{-0.005}$ & $0.871^{+0.017}_{-0.020}$ & $0.877^{+0.009}_{-0.006}$\\
        & $T$ (eV) 
        & $4.9^{+1.5}_{-0.9}$ & $3.1^{+0.9}_{-1.4}$ & $3.6^{+2.7}_{-2.9}$ & $4.0^{+0.9}_{-0.7}$ & $4.2^{+1.6}_{-0.9}$ & $4.8^{+11}_{-3.4}$ & $4.0^{+2.3}_{-0.8}$\\
    \hline
    $C$/d.o.f             & & 427/373 & 397/366 & 423/373 & 395/373 & 421/377 & 451/369 & 537/365\\
    Expected $C$-values     & & $399\pm28$ & $399\pm28$ & $399\pm28$ & $399\pm28$ & $399\pm28$ & $399\pm28$ & $399\pm28$ \\
\enddata
\tablecomments{The covering factor of the absorbers is fixed at $f_{\rm cov} = 0.5$, in consistent with Paper~I. Some values are at the imposed hard limit if it has errors of zero.\\
$^-$ this parameter is frozen \\
$^*$ This component is not included in the final model\\
}
\end{deluxetable*}

\begin{appendix}

\section{Calculation of Akaike information criterion and Detection Significance} \label{adx:DS}
Assessing the statistical significance of photoionized wind components is more complicated than assessing the significance of isolated Gaussian lines. A single absorption or emission line can often be characterized by the ratio of its best-fit normalization to its \(1\sigma\) uncertainty. In contrast, each ``$pion$" wind component produces a set of absorption and/or emission features whose strengths depend jointly on column density, ionization parameter, velocity, turbulent broadening, and the incident spectral energy distribution. Moreover, different wind components can partially overlap in energy, so that the apparent significance of one component may depend on how the neighboring components adjust during the fit.

For this reason, we do not rely only on the ratio \(N_{\rm H}/\sigma_{N_{\rm H}}\), although this quantity remains a useful diagnostic of whether a component is well constrained. Instead, we estimate the detection significance of each wind component using a conservative component-removal procedure. Starting from the final best-fit model, we remove one wind component at a time, refit the remaining continuum and wind parameters, and then compare the resulting fit statistic with that of the full model. This procedure allows the remaining model components to compensate as much as possible for the removed zone, and therefore gives a more conservative estimate of whether that component is independently required by the data.

We quantify the improvement in two complementary ways. First, we compute the change in Cash statistic, \(\Delta C\), between the model without the component and the final model. We then convert the corresponding \(p\)-value into a Gaussian-equivalent detection significance, using the number of additional free parameters associated with the removed component as the change in degrees of freedom. This value is reported as the detection significance, D.S. This approach is commonly used for nested spectral-model comparisons and provides an intuitive estimate of the significance in units of Gaussian \(\sigma\).

Second, we compute the Small Sample Akaike Information Criterion (AIC; \citealt{Akaike_1974, Emmanoulopoulos_2016}) as an independent model-selection diagnostic:
\[
{\rm AIC} = 2p - 2C_L + C + \frac{2p(p+1)}{n-p-1},
\]
where \(C\) is the C-statistic of the model being tested, \(C_L\) is the C-statistic of the final full model, \(n\) is the number of spectral bins, and \(p\) is the number of free parameters in the tested model. The final term is the small-sample correction, which penalizes models with more free parameters more strongly when \(n\) is not much larger than \(p\). Lower AIC values indicate a preferred model after accounting for model complexity. We define
\[
\Delta{\rm AIC} = {\rm AIC}_{\rm full} - {\rm AIC}_{\rm removed},
\]
so that negative values indicate that the full model, including the tested component, is preferred. As a guideline, \(|\Delta{\rm AIC}|<2\) indicates that the two models are statistically comparable, \(|\Delta{\rm AIC}|>2\) indicates substantial support for the model with the lower AIC, and \(|\Delta{\rm AIC}|>10\) indicates strong support \citep{Burnham_2002}. Therefore, components with \(\Delta{\rm AIC}<-10\) are strongly required, while those with \(-10<\Delta{\rm AIC}<-2\) have substantial support. Components with \(-2<\Delta{\rm AIC}<2\) are treated as marginal, and components with \(\Delta{\rm AIC}>2\) are not required.

In this work, we regard a wind component as significant when it has D.S. \(>3\sigma\) and is also supported by the AIC comparison. Components with \(1\sigma<{\rm D.S.}<3\sigma\) are reported as marginal when they are useful for tracking possible trends across event types, but they are not used for the primary physical interpretation or summed wind quantities. Components with D.S. \(<1\sigma\), or for which the column density regresses to zero, are excluded from the final model.

This procedure is more conservative than simply quoting the changes of C-stats after the addition of a component. By removing each component and refitting the remaining model, the test accounts for degeneracies among nearby wind zones and asks whether the data still require that component after the rest of the model has been allowed to respond. The best-fit parameters, D.S. values, and \(\Delta{\rm AIC}\) values for each component are listed in Table~\ref{table:parameters}.

\section{Calculation of CINDICITY} 
\label{adx:CIND}
The Color Intensity Index, or CINDICITY, is a new dimensionless parameter derived from the Hardness-Intensity Diagram (HID) in Figure \ref{fig:HRI} and \ref{fig:HID_cind}. This parameter captures the maximum variance in the joint space of count rate and hardness ratio. We compute the CINDICITY using the light curves extracted for each observation. For each time bin, we measure the full band (2.4--17.4 keV) count rate (R) and the ratio of the hard band (8.8--11.8) to soft band (3.0--4.0) count rate (H/S). We first transform the quantities in logarithmic space to linearize the multiplicative factors, such that
\begin{equation}
    x = \log_{10} R,
\end{equation}
\begin{equation}
    y = \log_{10} (H/S),
\end{equation}

,where $R$ is the count rate and $H/S$ is the hardness ratio. 

The two quantities are then standardized as
\begin{equation}
    x_s = \frac{x - \mu_x}{\sigma_x},
\end{equation}
\begin{equation}
    y_s = \frac{y - \mu_y}{\sigma_y},
\end{equation}
where $\mu_x$ and $\mu_y$ are the means of $\log_{10} R$ and $\log_{10}(H/S)$, and $\sigma_x$ and $\sigma_y$ are their corresponding standard deviations, computed over the full dataset. In this way, the count rate and hardness ratio are placed on the same normalized scale before performing PCA.

We then define the standardized data vector for each time bin as
\begin{equation}
    \mathbf{P} =
    \begin{pmatrix}
        x_s \\
        y_s
    \end{pmatrix}.
\end{equation}
Because the standardized cloud of points may not be centered exactly at the origin, we subtract the mean standardized position,
\begin{equation}
    \mathbf{P}_{\rm mean} =
    \left\langle \mathbf{P} \right\rangle,
\end{equation}
and construct the centered standardized vector
\begin{equation}
    \mathbf{P}_0 = \mathbf{P} - \mathbf{P}_{\rm mean}.
\end{equation}

We calculate the first principal component ($\textbf{PC1}$) using the singular value decomposition package (SVD; $numpy.linalg.svd$ \cite{harris_2020}) on the centered standardized points (\textbf{P}$_0$) in the reference of the mean ($P_0 = P - P_{mean}$), where $P = (x,y) =  (x_s(R), x_s(H/S))$. The sign of the PC1 vector is chosen such that increasing CINDICITY corresponds to the direction of softer and brighter states. 

The CINDICITY of each time bin is then defined as the scalar projection of $\mathbf{P}_0$ onto this axis:
\begin{equation}
    \mathrm{CINDICITY} = \mathbf{P}_0 \cdot \mathbf{PC1}.
\end{equation}
Equivalently, writing
\begin{equation}
    \mathbf{PC1} =
    \begin{pmatrix}
        p_x \\
        p_y
    \end{pmatrix},
\end{equation}
the CINDICITY can be expressed explicitly as
\begin{equation}
    \mathrm{CINDICITY} =
    (x_s - \langle x_s \rangle)\, p_x +
    (y_s - \langle y_s \rangle)\, p_y.
\end{equation}
By construction, CINDICITY is therefore a normalized one-dimensional coordinate that measures the location of each time bin along the dominant variability track in the HID, with positive values corresponding to softer, brighter states and negative values corresponding to harder, fainter states.

For each combined event type and selected spectral state (Flare, Postflare, After, Hard-dip, HS, LS, and LH), we compute the average count rate and hardness ratio, $\langle R \rangle$ and $\langle H/S \rangle$. These mean values are then transformed and standardized using the same logarithmic scaling and global normalization derived from the full time-resolved dataset. The resulting standardized vectors are centered using the same reference point and projected onto the principal axis $\mathbf{PC1}$. This ensures that the event-averaged CINDICITY values are defined consistently with the time-resolved measurements, with the location along the same dominant variability track in the HID.

\begin{figure*}
    \centering
    \includegraphics[width=1.0\linewidth]{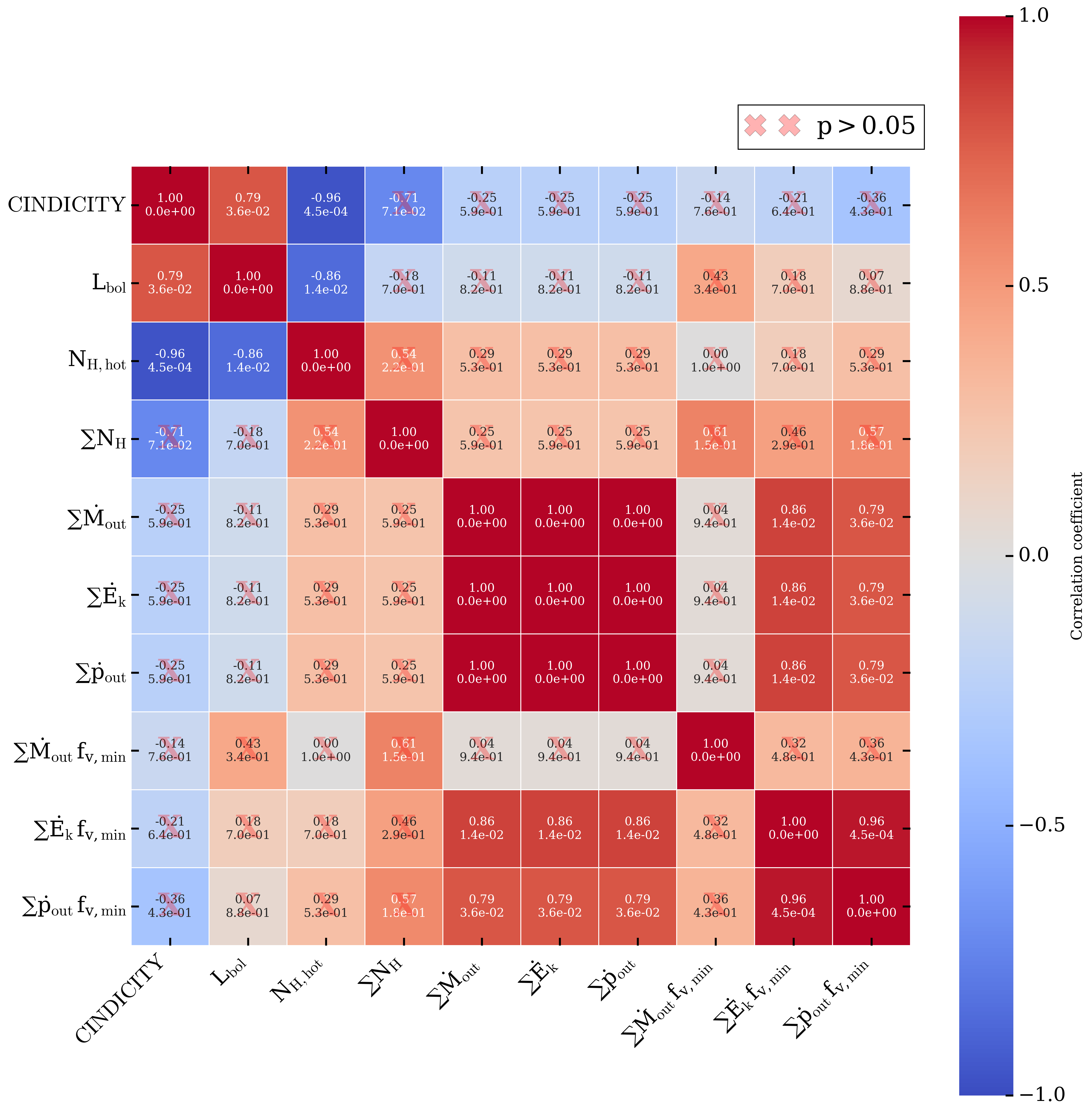}
    \caption{Spearman rank-correlation matrix for the integrated quantities of the seven time-selected and HID-selected spectra analyzed in this work.  Wind properties are calculated from significant components (D.S > $3 \sigma$). Each cell lists the correlation coefficient $\rho$ (upper value) and the corresponding two-sided $p$-value (lower value). Red crosses mark are for $p>0.05$. The variables include CINDICITY, $L_{\rm bol}$, the intrinsic obscurer column density $N_{\rm H,hot}$, the total wind column density $\Sigma N_{\rm H}$, the summed mass outflow rate $\Sigma \dot{M}_{\rm out}$, kinetic power $\Sigma \dot{E}_{\rm k}$, momentum rate $\Sigma \dot{p}_{\rm out}$, and the corresponding quantities after applying the minimum filling-factor correction.}
    \label{fig:correlation2}
\end{figure*}

\begin{figure*}
    \centering
    \includegraphics[width=1.0\linewidth]{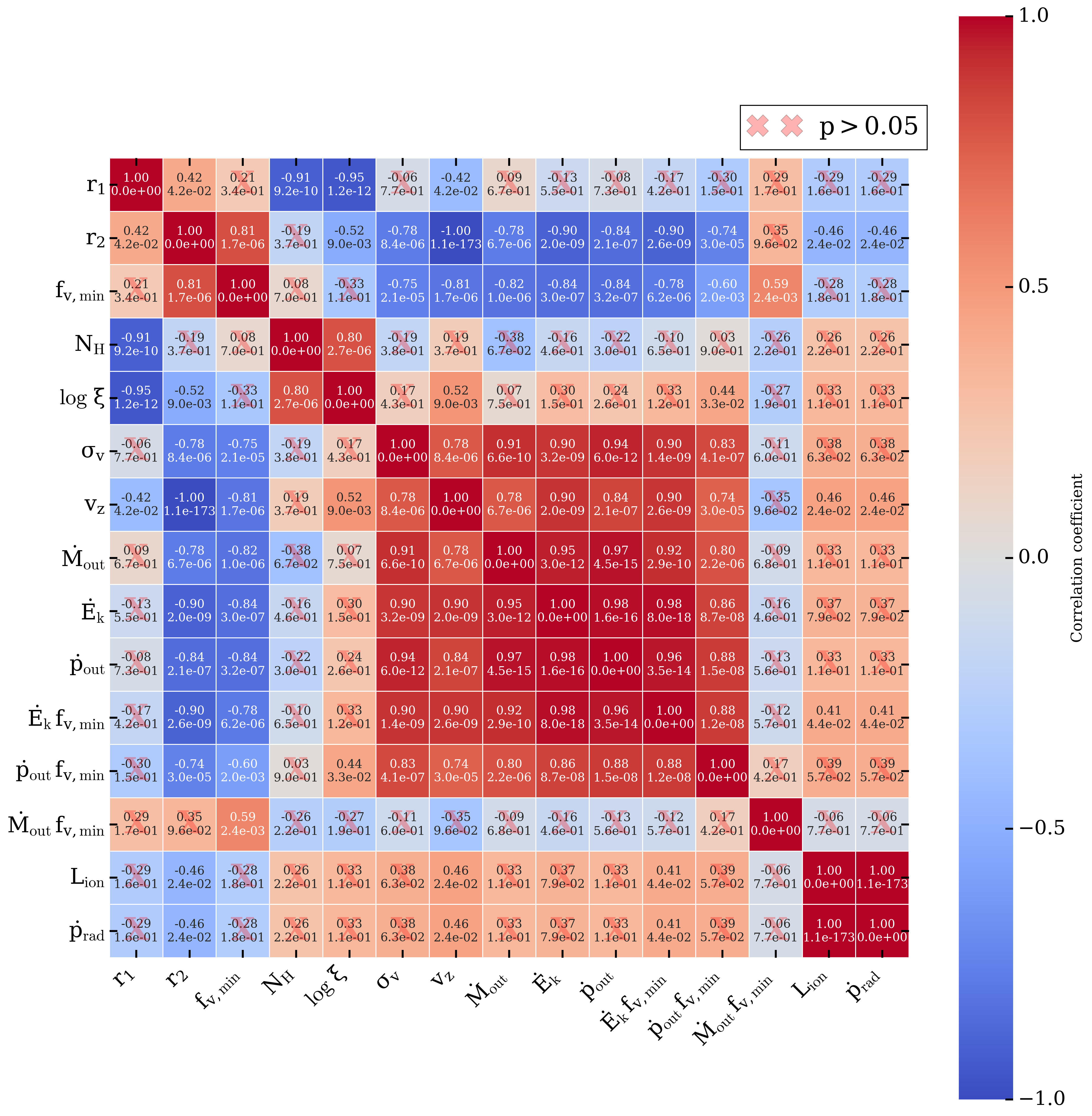}
    \caption{Spearman rank-correlation matrix for all significant (D.S > 3 $\sigma$) individual absorption components in seven time-selected and HID-selected spectra analyzed in this work. Each cell lists the correlation coefficient $\rho$ (upper value) and the corresponding two-sided $p$-value (lower value). Red crosses are for $p>0.05$. The variables include the radius scales $r_1$ and $r_2$, the minimum filling factor $f_{v,\min}$, the fitted wind parameters $N_{\rm H}$, $\log\xi$, $\sigma_v$, and $v_z$, the derived mass outflow rate $\dot{M}_{\rm out}$, kinetic power $\dot{E}_{\rm k}$, momentum rate $\dot{p}_{\rm out}$, the filling-factor-corrected energetics, and the incident ionizing luminosity $L_{\rm ion}$ and radiation momentum flux $\dot{p}_{\rm rad}$.}
    \label{fig:correlation1}
\end{figure*}

\section{Systematic Spearman's correlation scan}
To identify the dominant relationships more systematically, we carried out an exploratory Spearman-rank correlation scan using the seven uniformly analyzed spectra from this work. Figure~\ref{fig:correlation2} shows the map for integrated quantities in each spectrum, with each cell annotated with $\rho$ (upper value) and $p$ (lower value). The strongest event-level trends are that CINDICITY is positively correlated with $L_{\rm bol}$ ($\rho = 0.79$, $p = 3.6\times10^{-2}$) and anti-correlated with the intrinsic obscurer column density $N_{\rm H,hot}$ ($\rho = -0.96$, $p = 4.5\times10^{-4}$), indicating that brighter, softer states coincide with weaker intrinsic obscuration.  The bolometric luminosity is also anti-correlated with $N_{\rm H,hot}$ ($\rho=-0.86$), $p=1.4\times10^{-2}$), supporting the same picture.

The total wind column density, \(\Sigma N_{\rm H}\), shows a suggestive anti-correlation with CINDICITY ($\rho=-0.71$, $p=7.1\times10^{-2}$), consistent with the tendency for harder, fainter spectra to host larger wind columns, although this trend is not formally significant in the seven-spectrum sample. The relation between $N_{\rm H,hot}$ and $\Sigma N_{\rm H}$ is positive but weak in this event/HID-selected sample ($\rho=0.54$, $p=0.22$). Interestingly, when the five PV observations from Paper~I are added only as an illustrative comparison sample, this connection becomes more suggestive ($\rho=0.56$, $p=5.8\times10^{-2}$), which strengthens the case that the intrinsic obscurer and total wind column may be physically connected.

The wind energetics are, as expected, tightly correlated among themselves. After applying the minimum filling-factor correction, $\Sigma \dot{E}_{\rm k} f_{v,\min}$ remains significantly correlated with $\Sigma \dot{p}_{\rm out} f_{v,\min}$, but neither quantity shows a significant dependence on CINDICITY. Thus, the strongest event-level connection is between the source spectral state and the intrinsic obscurer, while the link between spectral state and global wind energetics appears weaker, with a marginal but suggestive connection between intrinsic obscuration and the total wind column. This dynamical connection hints that both may respond to changes in the central engine and be shaped by related physical or geometric changes in the inner source.
 
Figure~\ref{fig:correlation1} shows the same analysis performed on all significant (D.S > 3 $\sigma$) individual absorption components and their derived quantities. Several expected structural relations are recovered. The ionization-based radius scale $r_1$ is strongly anti-correlated with both $N_{\rm H}$ and $\log\xi$, while the escape-based radius scale $r_2$ is strongly anti-correlated with $v_z$, $\sigma_v$, and the derived energetics. The minimum filling factor $f_{v,\min}$ is also strongly anti-correlated with $v_z$, $\sigma_v$, $\dot{M}_{\rm out}$, $\dot{E}_{\rm k}$, and $\dot{p}_{\rm out}$, reflecting the increasingly clumpy nature of the faster outflows. Among the directly fitted parameters, $N_{\rm H}$ and $\log\xi$ are positively correlated ($\rho = 0.80$, $p = 2.7\times10^{-6}$), consistent with the AMD trend discussed below. The wind energetics correlate most strongly with velocity and turbulent width, indicating that the fastest and broadest components dominate the feedback budget. By contrast, $L_{\rm ion}$ and $\dot{p}_{\rm rad}$ show only weak to moderate correlations with the derived energetics, suggesting that the instantaneous ionizing luminosity alone is not the sole driver of the observed wind diversity.

\end{appendix}

\begin{acknowledgments}
We thank the anonymous referee for thoughtful and constructive comments that helped improve the statistical tests of wind variability and clarify the interpretation of the absorber radius constraints.
\end{acknowledgments}

\bibliography{main}{}
\bibliographystyle{aasjournal}

\end{document}